\documentclass[journal]{IEEEtran}
\usepackage{ifpdf}
\usepackage{cite}
\usepackage[cmex10]{amsmath}
\usepackage{amssymb}
\usepackage{algorithm}
\usepackage{algorithmic}
\usepackage{array}
\usepackage{mdwmath}
\usepackage{mdwtab}
\usepackage{eqparbox}
\usepackage{subfig}
\usepackage{stfloats}
\usepackage{alphalph}
\usepackage{multirow}
\usepackage{url}
\usepackage{setspace}
\usepackage{graphicx}
\usepackage{slashbox}

\newcommand{\PreserveBackslash}[1]{\let\temp=\\#1\let\\=\temp}
\newcolumntype{C}[1]{>{\PreserveBackslash\centering}p{#1}}
\newcolumntype{R}[1]{>{\PreserveBackslash\raggedleft}p{#1}}
\newcolumntype{L}[1]{>{\PreserveBackslash\raggedright}p{#1}}

\DeclareGraphicsExtensions{.pdf}

\begin{document}
%
% paper title
% Titles are generally capitalized except for words such as a, an, and, as,
% at, but, by, for, in, nor, of, on, or, the, to and up, which are usually
% not capitalized unless they are the first or last word of the title.
% Linebreaks \\ can be used within to get better formatting as desired.
% Do not put math or special symbols in the title.
\title{DeepLofargram: A Deep Learning based Fluctuating Dim Frequency Line Detection and Recovery
\thanks{This work was supported in part by the National Key R\&D Program of China under Grant 2016YFC1400200, in part by the National Natural Science Foundation of China under Grant 61671388, and in part by the National Engineering Laboratory for Integrated Aero-Space-Ground-Ocean Big Data Application Technology. (\emph{Corresponding author: Yina Han and Qingyu Liu})}}
%
%
% author names and IEEE memberships
% note positions of commas and nonbreaking spaces ( ~ ) LaTeX will not break
% a structure at a ~ so this keeps an author's name from being broken across
% two lines.
% use \thanks{} to gain access to the first footnote area
% a separate \thanks must be used for each paragraph as LaTeX2e's \thanks
% was not built to handle multiple paragraphs
%

\author{Yina~Han,~\IEEEmembership{Member,~IEEE,}
        Yuyan~Li,
        %Jianjing~Deng,
        Qingyu~Liu,
        and~Yuanliang~Ma% <-this % stops a space
\thanks{Yina~Han, Yuyan~Li and Yuanliang~Ma are with the School of Marine Science and Technology, Northwestern Polytechnical University, Xi'an 710072, China, and also with the Key Laboratory of Ocean Acoustics and Sensing, Ministry of Industry and Information Technology, Xi'an 710072, China. (E-mail: \{yina.han, ylma\}@nwpu.edu.cn)}% <-this % stops a space
\thanks{Qingyu~Liu is with the Institute of Navy Research of China, Beijing 100073, China (E-mail: qyliu@nwpu.edu.cn).}}% <-this % stops a space

% note the % following the last \IEEEmembership and also \thanks -
% these prevent an unwanted space from occurring between the last author name
% and the end of the author line. i.e., if you had this:
%
% \author{....lastname \thanks{...} \thanks{...} }
%                     ^------------^------------^----Do not want these spaces!
%
% a space would be appended to the last name and could cause every name on that
% line to be shifted left slightly. This is one of those "LaTeX things". For
% instance, "\textbf{A} \textbf{B}" will typeset as "A B" not "AB". To get
% "AB" then you have to do: "\textbf{A}\textbf{B}"
% \thanks is no different in this regard, so shield the last } of each \thanks
% that ends a line with a % and do not let a space in before the next \thanks.
% Spaces after \IEEEmembership other than the last one are OK (and needed) as
% you are supposed to have spaces between the names. For what it is worth,
% this is a minor point as most people would not even notice if the said evil
% space somehow managed to creep in.

% The paper headers
\markboth{Journal of \LaTeX\ Class Files,~Vol.~14, No.~8, August~2015}%
{Shell \MakeLowercase{\textit{et al.}}: Bare Demo of IEEEtran.cls for IEEE Journals}
% The only time the second header will appear is for the odd numbered pages
% after the title page when using the twoside option.
%
% *** Note that you probably will NOT want to include the author's ***
% *** name in the headers of peer review papers.                   ***
% You can use \ifCLASSOPTIONpeerreview for conditional compilation here if
% you desire.

% If you want to put a publisher's ID mark on the page you can do it like
% this:
%\IEEEpubid{0000--0000/00\$00.00~\copyright~2015 IEEE}
% Remember, if you use this you must call \IEEEpubidadjcol in the second
% column for its text to clear the IEEEpubid mark.

% use for special paper notices
%\IEEEspecialpapernotice{(Invited Paper)}

% make the title area
\maketitle

% As a general rule, do not put math, special symbols or citations
% in the abstract or keywords.
\begin{abstract}
This paper investigates the problem of dim frequency line detection and recovery in the so-called lofargram. Theoretically, time integration long enough can always enhance the detection characteristic. But this does not hold for irregularly fluctuating lines. Deep learning has been shown to perform very well for sophisticated visual inference tasks. With the composition of multiple processing layers, very complex high level representation that amplify the important aspects of input  while suppresses irrelevant variations can be learned. Hence we propose a new DeepLofargram, composed of deep convolutional neural network and its visualization counterpart. Plugging into specifically designed multi-task loss, an end-to-end training jointly learns to detect and recover the spatial location of potential lines. Leveraging on this deep architecture, the performance boundary is $-24$dB on average, and $-26$dB for some. This is far beyond the perception of human visual and significantly improves the state-of-the-art.
\end{abstract}

% Note that keywords are not normally used for peerreview papers.
\begin{IEEEkeywords}
deep learning, lofargram, frequency line detection.
\end{IEEEkeywords}

% For peer review papers, you can put extra information on the cover
% page as needed:
% \ifCLASSOPTIONpeerreview
% \begin{center} \bfseries EDICS Category: 3-BBND \end{center}
% \fi
%
% For peerreview papers, this IEEEtran command inserts a page break and
% creates the second title. It will be ignored for other modes.
\IEEEpeerreviewmaketitle

\section{Introduction}
\label{sec:In}

\IEEEPARstart{D}{etecting} the presence of quiet man-made targets in passive sonar systems remains one of the most challenging and important problems in marine surveillance\cite{HayesSMBE2013}. In general, due to the rotating pieces of a motor, acoustic wave with a certain tone is emitted, and  a frequency versus time image called lofargram (low frequency analysis and recording) is constructed to help to detect such emission, usually in the form of a frequency line~\cite{Maskell2001}.

%A lofargram is a gray-level image representation of this acoustic emission's periodogram, and the role of target detection in the lofargram is to resolve the coherent pixels of the this frequency line from ambient noise~\cite{}.

%Generally, remote targets correspond to weak periodic phenomena submerged in the ambient noise, namely low Signal-to-Noise Ratio (SNR) in a spectrogram. Moreover, due to the nature of the observed phenomenon and the nature of the signals of interest, the structure of the frequency line may vary with time, such as vertical straight line, oblique straight line, and the most challenging irregular fluctuating line~\cite{}.  Theoretically, time integration of the lofargram long enough can always enhance the detection characteristic, given that the energy accumulation for target is  greater than for surrounding clutters. But this does not hold for irregularly fluctuating frequency lines.

% 就从频率变化快慢和信噪比两个角度来综述方法，这也是本文主要展开的视角
This problem can be traced back to the mid 1940s by Koenig et al.~\cite{KoenigDL1946}, and has attracted great interest from a variety of backgrounds, ranging from image processing~\cite{AbelLL1992,DiT1996,ThomasS2013}, neural networks~\cite{KendallHN1993}, statistical modelling~\cite{RifeB1974,ScharfE1981,ParisJ2001,ParisJ2003}, etc.  Nevertheless the low signal-to-noise ratio (SNR) in remote and quiet sensing applications compound with the variability of the frequency line caused by the nature of the observed target is still of great challenge~\cite{LiLY2008,ThomasS2013}. Specifically, for dim but vertical straight lines, time integration long enough can always enhance the detection characteristic. But this does not hold for irregularly fluctuating frequency line~\cite{LiLY2008}. Sophisticated image processing and neural network based methods can handle complex visual patterns, but become unreliable in low SNRs~\cite{LampertO2011}. The idea of statistical models is to model the fluctuation of the potential line using statistical priori and then conduct a probabilistic integration over the spectral power~\cite{ParisJ2001,ParisJ2003}. For a more comprehensive review please refer to~\cite{LampertO2010}. The reported state-of-the-art methods are  contour energy minimisation~\cite{ThomasS2013} for low line shape variation, and Hidden Markov Models~\cite{ParisJ2003} for fluctuating line shape variation.

Recently, deep convolutional neural networks (ConvNet) have led to a series of breakthroughs for image classification~\cite{krizhevsky2012,SimonyanZ2014a,YangDZYL2019,XuLYDL2019}. There is broad consensus that the success of deep ConvNet results from its deep representation power, by which the raw input are transformed to abstract high level features that amplifies the important aspects  while suppresses irrelevant variations~\cite{LeCunBH2015}. Encouraged by this success, various methods, such as deconvolutional network~\cite{ZeilerF2014}, data gradient~\cite{SimonyanVZ2014} and guided backpropagation~\cite{SpringenbergDBR2015}, are developed to analyze the aspects of visual appearance captured inside a deep model. This further advances the progress of structured prediction of the image~\cite{LongSD2015}.

To draw a parallel, the main challenge of visual inference tasks in computer vision is to reveal the ``semantic consistency'' from diverse visual appearances. While the main challenge in lofargram can be deemed as resolving the ``spatial coherency'' of potential frequency line from overwhelming ambient noise. Hence we argue that the use of high level abstract deep ConvNet architecture, as semantic deep representation in computer vision,  may offer another promising venue to empower a lofargram to tackle visually invisible and irregularly fluctuating frequency lines.

In this letter, we propose a ``DeepLofargram'' framework, by which we first leverage on the power of deep ConvNet for image classification, and conduct a binary hypothesis testing on noise dominating lofargram; then inspired by the idea of deep network visualization, spatial support
pixels for the positive hypothesis captured by the deep ConvNet are coarsely output as a frequency line saliency map. Plugging specifically designed multi-task loss an end-to-end training can be effectively conducted. We validate our approach on representative simulation datasets, where qualitative results, ablations, and intuitive examples demonstrate our method's ability to detect and recovery fluctuating dim frequency lines.

\section{System Architecture and Training}

\begin{figure}
\centering
\includegraphics[width=0.4\textwidth]{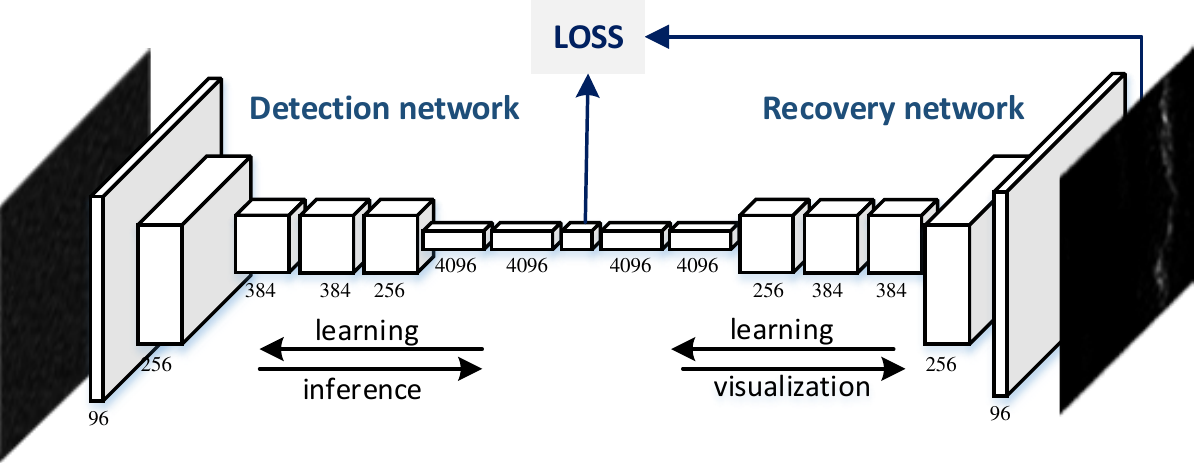}
\caption{Framework of DeepLofargram.}
\label{fig:framework}
\end{figure}

Fig.~\ref{fig:framework} illustrates the DeepLofargram architecture. A DeepLofargram network is composed of two modules - frequency line detection and frequency line recovery networks. The detection network corresponds to several convolutional layers to extract high level deep features, followed by $3$ fully connected layers to conduct a binary hypothesis testing,
\begin{equation}
\begin{split}
H_0: &\  \mathrm{noise\ only}\\
H_1: &\  \mathrm{frequency\ line\ present}+\mathrm{noise}
\end{split}
\end{equation}
Whereas the recovery network is its visualisation counterpart that numerically generates an image with the specific spatial support of the frequency line captured by the $H_1$ hypothesis of the deep ConvNet.

We employ AlexNet~\cite{krizhevsky2012} and VggNet~\cite{SimonyanZ2014a} for frequency line detection with its last $1000$-way classification layer (for the $1000$ categories respectively) replaced with a $2$-way classification layer (for $H_0$ and $H_1$ respectively). Starting with the raw lofargrams, it composes $5$ convolutional layers altogether with non-linear rectification and pooling operations performed between them, and $3$ fully connected layers that outputs softmax probability estimates over the $2$ hypothesis. When $H_1$ hypothesis is declared by the detection network, the follow-up recovery network is activated. Our recovery network is a mirror version of the detection network and generates an image that maximises the $H_1$ score, which is expected to be the dim frequency line captured by the deep network. Contrary to detection module that maps a lofargram to $2$ detection scores through feedforwarding, recovery module computes the frequency line saliency map using the derivative of $H_1$ hypothesis score with respect to the lofargram. More details are described in the following subsections.

\subsection{Frequency Line Saliency Map}
Given a lofargram $I\in [0,255]^{T\times K}$, where $T$ and $K$ represent the number of time slots and frequency bins respectively, and a detection network with the $2$ output scores $s_0$ and $s_1$ for hyperthesis $H_0$ and $H_1$ respectively, we can expect that the  frequency line reside in the part of the lofargram that most activates the $s_1$ score. This is referred to as frequency line saliency map $S\in [0,1]^{T\times K}$, and can be achieved by computing the derivative of $s_1$ with respect to the lofargram $I$ to indicate pixels that most affect the class score~\cite{SimonyanVZ2014}. Hence we invert the data flow of the detection network, going from score activation layer down to the lofargram through the gradient back-propagation of RELU rectification, fully connected, max-pooling and convolutional layers.

Specifically, for the RELU rectification layer $X_{n+1}=\max(X_n,0)$, the sub-gradient takes the form~\cite{SimonyanVZ2014} $\frac{\partial s_1}{\partial X_n}=\frac{\partial s_1}{\partial X_{n+1}}\cdot\mathbf{1}(X_n>0)$. While in deconvolutional network~\cite{ZeilerF2014} RELU layer reconstruction is equivalent to $\frac{\partial s_1}{\partial X_n}=\frac{\partial s_1}{\partial X_{n+1}}\cdot\mathbf{1}(X_{n+1}>0)$. Combining these two together, a guided version of~\cite{SimonyanVZ2014} is proposed by~\cite{SpringenbergDBR2015}: $\frac{\partial s_1}{\partial X_n}=\frac{\partial s_1}{\partial X_{n+1}}\cdot\mathbf{1}(X_n>0)\cdot\mathbf{1}(X_{n+1}>0)$. For the fully connected layer $X_{n+1}=W^T X_n$, the gradient is computed as $\frac{\partial s_1}{\partial X_n}=\frac{\partial s_1}{\partial X_{n+1}}\cdot W$, where $W$ is the weight matrix of this layer. For the max-pooling layer $X_{n+1}(p)=\max_{q\in\Omega(p)}X_n(q)$, where the element $p$ of the output feature map is computed by pooling over the corresponding spatial neighbourhood $\Omega(p)$ of the input. The sub-gradient is computed as $\frac{\partial s_1}{\partial X_n(s)}=\frac{\partial s_1}{\partial X_{n+1}(p)}\cdot\mathbf{1}(s=\mathrm{argmax}_{q\in\Omega(p)}X_n(q))$, where $\mathbf{1}$ is the element-wise indicator function. Finally, consider the convolutional layer $X_{n+1}=X_n\ast K_n$, the gradient is computed as $\frac{\partial s_1}{\partial X_n}=\frac{\partial s_1}{\partial X_{n+1}}\ast \widehat{K_n}$, where $K_n$ and $\widehat{K_n}$ are the convolution kernel and its flipped version, respectively.

\subsection{Loss Function}
A DeepLofargram network has two output layers associated with the detection and recovery modules respectively. The first outputs a discrete probability distribution (per lofargram), $p = (p_0, p_1)$, over the $2$ hypothesis $H_0$ and $H_1$. As usual, $p$ is computed by a softmax over the $2$ output scores $s_h$ of a fully connected layer: $p_h=\frac{\exp s_h}{\exp s_0 + \exp s_1}$. The second outputs the probability map $P\in [0,1]^{T\times K}$ over the lofargram that indicating where the frequency line is for $H_1$ hypothesis.

Each training lofargram is labeled with a ground-truth hypothesis $h\in\{0,1\}$ and a ground-truth frequency line map $F\in\{0,1\}^{T\times K}$.  By convention the $H_0$ hypothesis is labeled $h = 0$ and  the $H_1$ hypothesis is labeled $h = 1$. We use a multi-task loss $L$ on each labeled lofargram to jointly train for detection and frequency line recovery:
\begin{equation}
L(p,h,P,F)=L_{\mathrm{det}}(p,h)+\mathbf{1}(h = 1)L_{\mathrm{rec}}(P,F)
\end{equation}
in which $L_{\mathrm{det}}(p, h) = -\log p_h$ is log loss for true hypothesis $h$. The second task loss $L_{\mathrm{rec}}(P,F)$ is defined over a ground-truth frequency line map $F$ and a predicted frequency line map $P$, again for $H_1$. $\mathbf{1}(h = 1)$ is an indicator function evaluating to $1$ when $h=1$ and $0$ otherwise. For $H_0$ there is no notion of a ground-truth frequency line and hence $L_{\mathrm{rec}}$ is ignored.

Given a training lofargram, $L_{\mathrm{rec}}(P,F)$ is computed over $P_{t,k}$ and $F_{t,k}$ for each $t=0,\dots,T-1$ and $k=0,\dots,K-1$. Typically, the distribution of frequency/non-frequency pixels is heavily biased: $90\%$ of the ground truth is non-frequency. To balance the loss between frequency/non frequency classes, we adopt the class-balanced cross-entropy loss function~\cite{},
\begin{equation}
L_{\mathrm{res}}(P,F)=-\beta\sum_{t,k\in F_{+}}\log P_{t,k}
-(1-\beta)\sum_{t,k\in F_{-}}\log (1-P_{t,k})
\end{equation}
where $\beta = |F_{-}|/|F|$ and $1-\beta = |F_{+}|/|F|$. $|F_{+}|$ and $|F_{-}|$ denote the frequency and non-frequency ground truth label sets, respectively. $P$ is computed using sigmoid function on the aforementioned frequency line saliency map $S$, that is $P_{t,k}=\frac{1}{1+e^{-S_{t,k}}}$, $t=0,\ldots,T-1,\ k=0,\ldots, K-1$.

%The hyper-parameter $\lambda$ in Eq.~\ref{} controls the balance between the two task losses. All experiments use $\lambda = 1$.

%
%Given A Training Lofargram, $L_{\Mathrm{Rec}}(P,F)$ Is Computed Over $P_{T,K}$ And $F_{T,K}$ For Each $T=0,\Dots,T-1$ And $K=0,\Dots,K-1$. Typically, The Distribution Of Frequency/Non-Frequency Pixels Is Heavily Biased: $90\%$ Of The Ground Truth Is Non-Frequency. To Balance The Loss Between Frequency/Non Frequency Classes, We Adopt The Class-Balanced Cross-Entropy Loss Function~\Cite{},
%\Begin{Equation}
%L_{\Mathrm{Res}}(P,F)=-\Beta\Sum_{T,K\In F_{+}}\Log P_{T,K}
%-(1-\Beta)\Sum_{T,K\In F_{-}}\Log (1-P_{T,K})
%\End{Equation}
%Where $\Beta = |F_{-}|/|F|$ And $1-\Beta = |F_{+}|/|F|$. $|F_{+}|$ And $|F_{-}|$ Denote The Frequency And Non-Frequency Ground Truth Label Sets, Respectively. $P$ Is Computed Using Sigmoid Function On The Aforementioned Frequency Line Saliency Map $S$, That Is $P_{T,K}=\Frac{1}{1+E^{-S_{T,K}}}$, $T=0,\Ldots,T-1,\ K=0,\Ldots, K-1$.

\subsection{Training Phase}

The training was conducted by optimising the above loss function using mini-batch gradient descent with Adam. The batch size was set to $128$. The training procedure was regularised by weight decay and dropout for the first two fully-connected layers as~\cite{krizhevsky2012,SimonyanZ2014a}. Here the $L_2$ penalty multiplier was set to $10^{-3}$ and the dropout ratio was set to $0.5$. Training deep network with low SNR lofargrams is difficult to converge. Hence we pre-trained our network with relatively high SNR lofargrams ($-20$dB and $-23$dB) using the learning rate of $10^{-3}$. Then we decreased the learning rate to $10^{-4}$ and trained our network with low SNR lofargrams ($-24$dB, $-25$dB and $-26$dB). Both of the above training procedures are terminated after $100K$ iterations ($55$ epochs).

%the entire number of weights in our net is not greater than the number of weights in AlexNet. To facilitate convergence, we initialize the network with detection loss only and lofargrams under relatively high SNR first, such as $-15$dB that can be perceived by human visual system. Then we fine-tune the pre-trained network with joint loss of detection and recovery in~\ref{} later.
%
%
%The training is carried out

%\begin{figure}
%\centering
%\subfloat[][$H_0$]{
%\includegraphics[width=0.13\textwidth]{noise2}}
%\subfloat[][$-5$dB]{
%\includegraphics[width=0.13\textwidth]{LF-5dB}
%\label{fig:groundtruth}}
%\subfloat[][$-22$dB]{
%\includegraphics[width=0.13\textwidth]{LF-22dB2}}
%%\subfloat[][$-23$dB]{
%%\includegraphics[width=0.13\textwidth]{LF-23dB2}}
%%\subfloat[][$-24$dB]{
%%\includegraphics[width=0.13\textwidth]{LF-24dB2}}
%%\subfloat[][$-25$dB]{
%%\includegraphics[width=0.13\textwidth]{LF-25dB2}}
%%\subfloat[][$-26$dB]{
%%\includegraphics[width=0.13\textwidth]{LF-26dB2}}
%\caption{Lofargram example: (2.1) noise only $H_0$ lofargram; (2.2)-(2.7) $H_1$ lofargrams in different SNRs.}
%\label{fig:dataset}
%\end{figure}

\section{Experiments}
This section first introduces utilized synthesized datasets and evaluation criteria. Then a number of ablation experiments analyzing the important aspects of our approach are performed. Comparisons with the state-of-the-art are reported last.

\subsection{Dataset}
To examine the capability of detecting irregularly fluctuating frequency lines in low SNRs, we adopt the synthesizing model proposed by Paris~\cite{ParisJ2003}:
\begin{equation}
s_t = a_t\sin(2\pi f_t+\phi_t)+\epsilon_t,\quad t=0,\ldots, T-1
\end{equation}
where the observation $s_t$ is assumed to be a sinusoid corrupted by an additive zero-mean Gaussian white noise $\epsilon_t\sim \mathcal{N}(0,\sigma^2_t)$, and the time behavior of instantaneous frequencies $\{f_t,t\in\{0,\ldots,T-1\}\}$ is modeled as a random walk. For fair comparison, a benchmark dataset composed of $1000$ noise-only lofargrams, and $1000$ lofargrams in specified SNR is synthesized. We split the dataset into $90\%$ training and $10\%$ testing for all the experiments. Specifically, we first generating $1000$ instances of Gaussian noises and $1000$ instances of sinusoids with random walk frequencies by Monte-Carlo simulation with the same setting as~\cite{ParisJ2003}. Then Gaussian noises are transformed into the frequency domain to form $H_0$ lofargrams. Gaussian noises added to assigned sinusoids with SNR defined in time domain $\mathrm{SNR}\stackrel{\Delta}{=}10\log_{10}\left(\frac{a^2_k}{2\sigma^2_k}\right)$ are transformed to form $H_1$ lofargrams. As shown in Fig.~\ref{fig:dataset}, when below $-20$dB, potential frequency lines are totally beyond the perception of human visual system, and this is the case of our focus.

\subsection{Evaluation Metrics}
We use $2$ standard metrics: ROC curve and line location accuracy~\cite{ThomasS2013} to evaluate the performance of detection and the accuracy of recovered  frequency line respectively. Given the continuous output scores $s_1$s over the test dataset,  decisions on $H_1$ hypothesis are made based on a threshold. Then the  probability of detection (PD) and false alarm rate (FAR) are computed as PD$=\frac{TP}{TP+FN}$ and FAR$=\frac{FP}{FP+TN}$, where $TP$, $FN$, $FP$ and $TN$ denotes the number of True Positives, False Negatives, False Positives and True Negatives. Changing the threshold continuously yields the ROC curve.

To further evaluate the quality of recovered frequency line, the line location accuracy ($LLA$) metric is used:
$LLA=\frac{1}{\max(|P|,|F|)}\sum_{(l,m)\in P}\frac{1}{1+\lambda\min_{(i,j)\in F}(\|[l,m]-[i,j]\|^2)}$, where $P$ and $F$ denote the predicted frequency line map and the ground truth. $|\cdot|$ accumulates the nonzero entries in a map and $\|[l,m]-[i,j]\|$ is the Euclidean distance between the recovered and actual line pixels. We set $\lambda=1$  as in~\cite{ThomasS2013}.

%\begin{figure}
%\centering
%\subfloat[][Precision-Recall Curve]{
%\includegraphics[width=0.24\textwidth]{PRC3}}
%\subfloat[][Line Location Accuracy]{
%\includegraphics[width=0.24\textwidth]{LLA}}
%\caption{The two evaluation metrics under varied low SNRs.}
%\label{fig:evametric}
%\end{figure}

\subsection{Ablation Analysis}

\subsubsection{Data Augmentation}

\begin{figure}
\centering
\subfloat[][Lofargram]{
\includegraphics[width=0.14\textwidth]{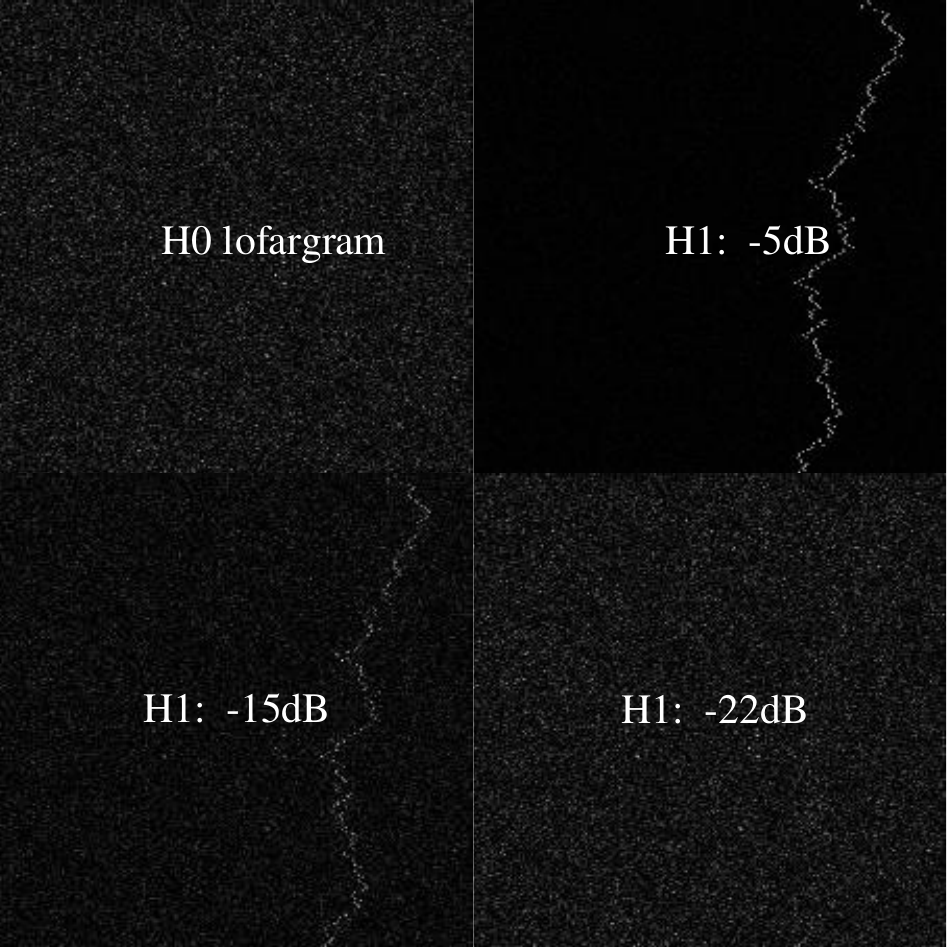}
\label{fig:dataset}}
\subfloat[][$-22$dB]{
\includegraphics[width=0.15\textwidth]{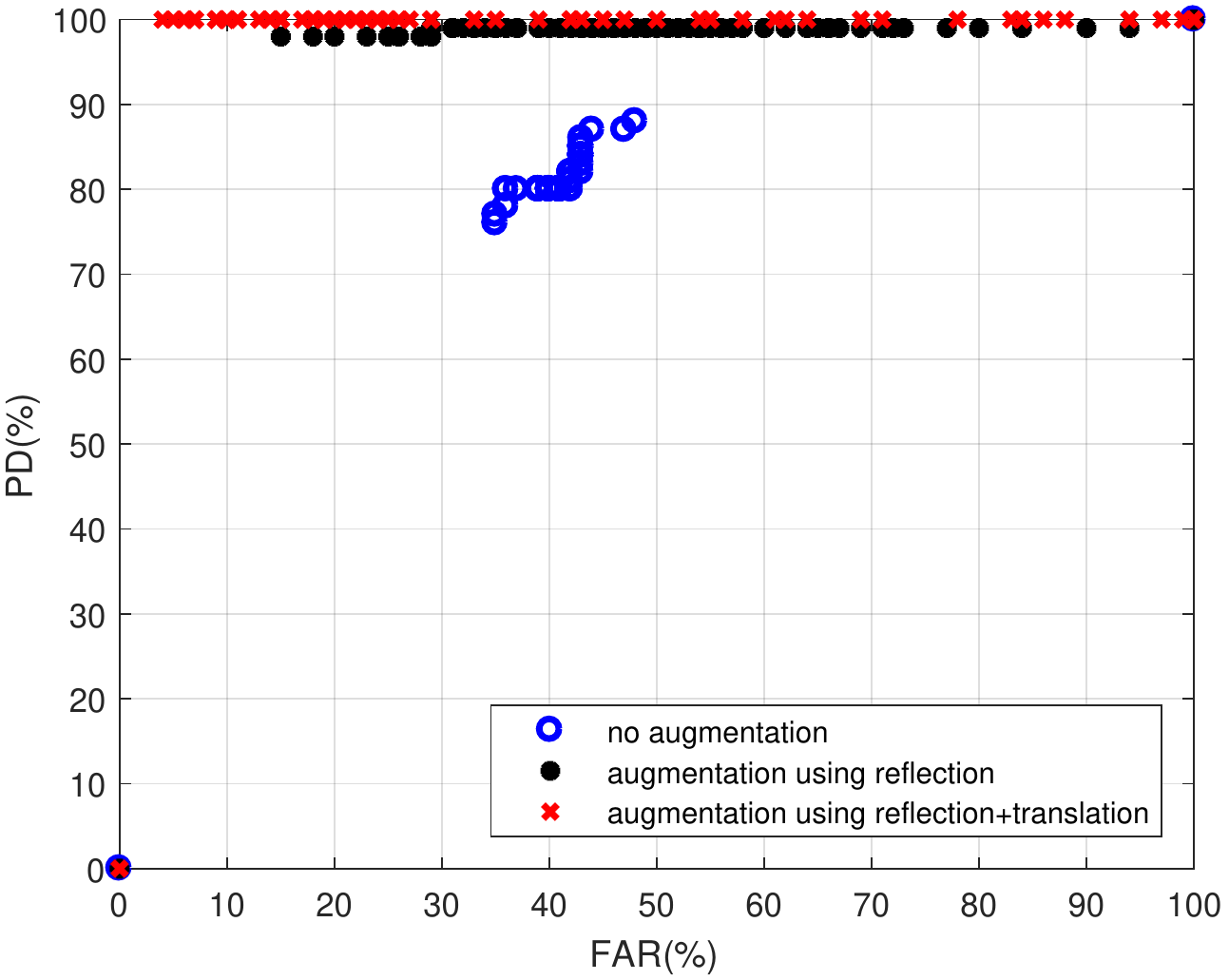}
\label{fig:augmentation-22}}
\subfloat[][$-23$dB]{
\includegraphics[width=0.15\textwidth]{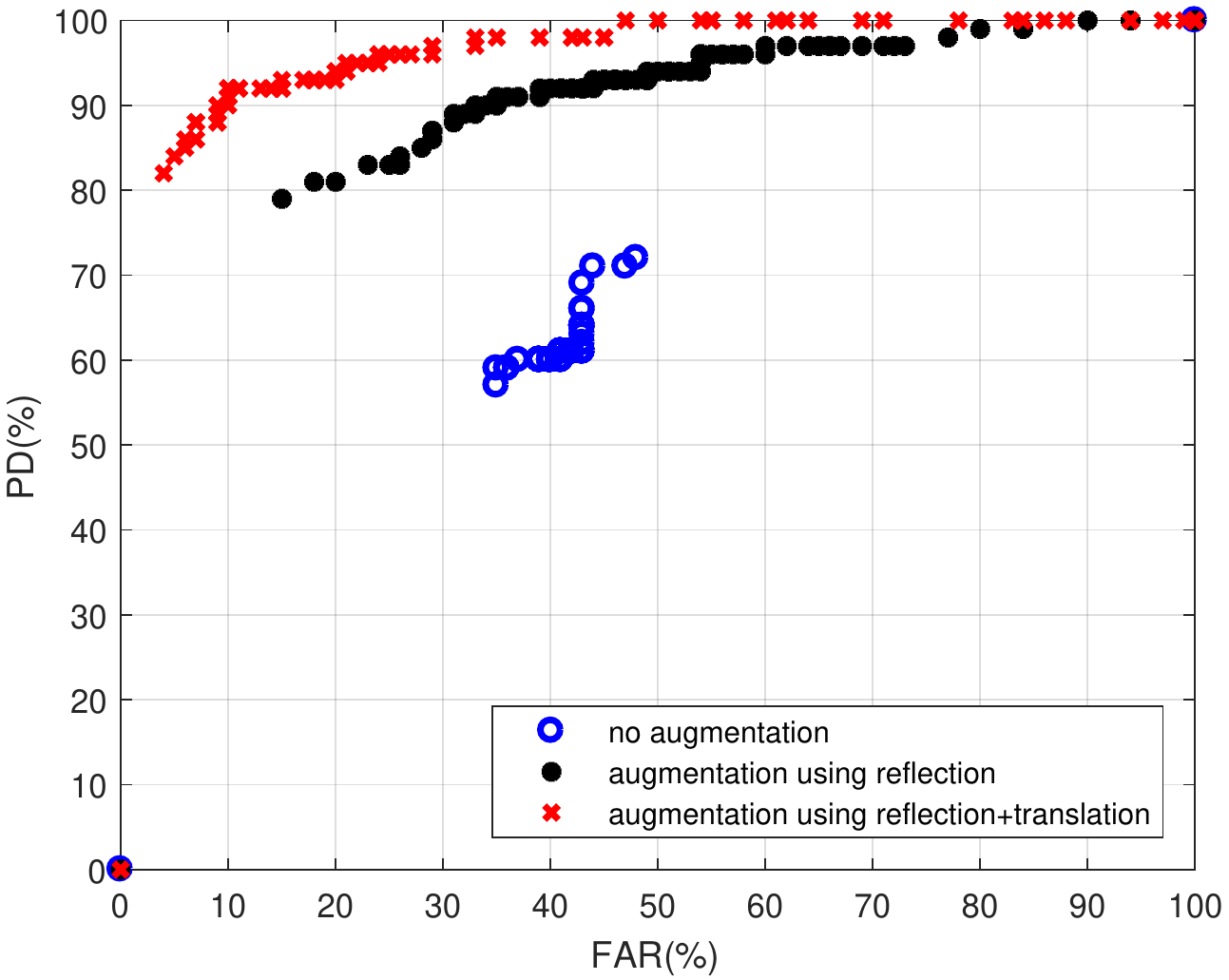}
\label{fig:augmentation-23}}\\
\subfloat[][$-24$dB]{
\includegraphics[width=0.15\textwidth]{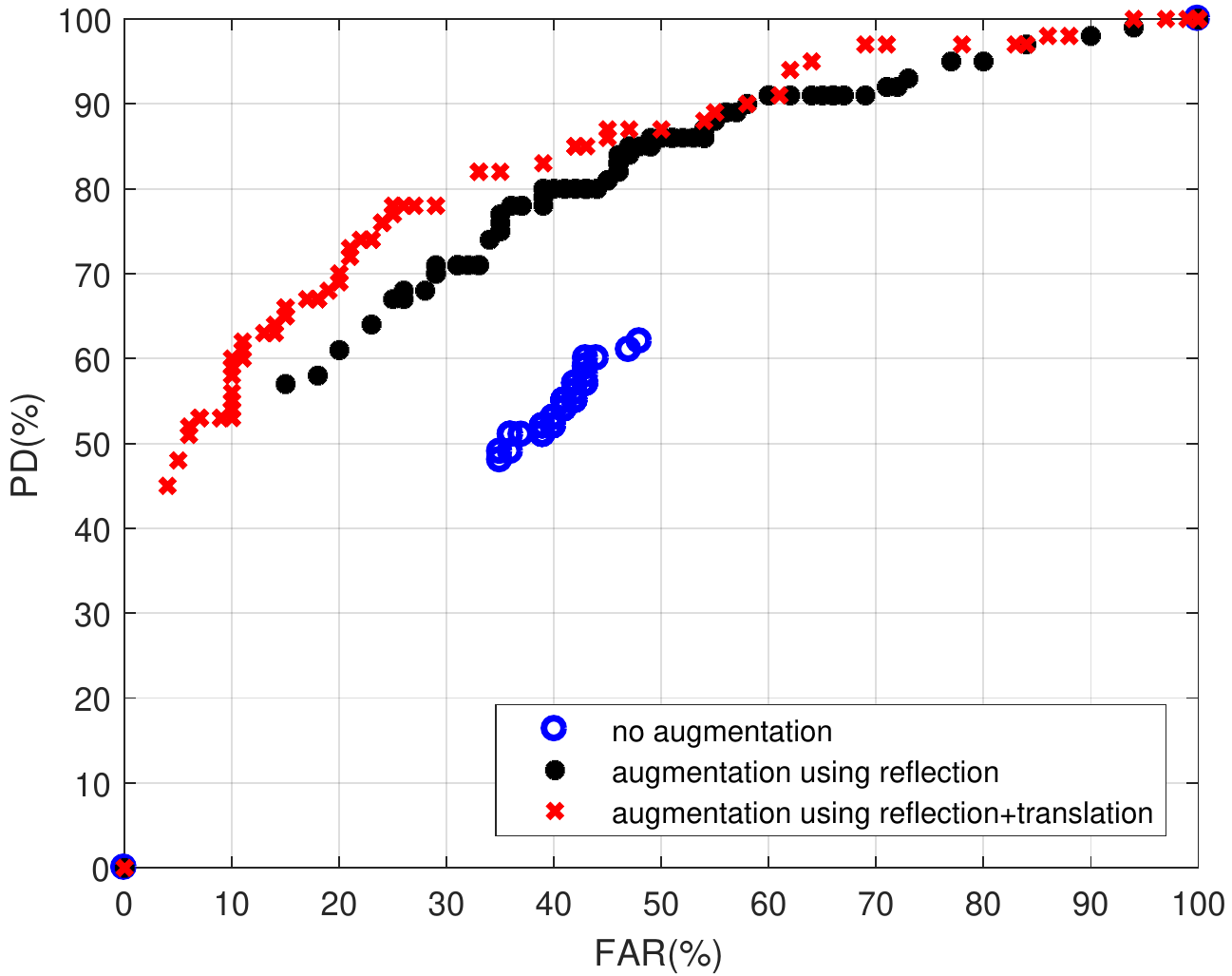}
\label{fig:augmentation-24}}
\subfloat[][$-25$dB]{
\includegraphics[width=0.15\textwidth]{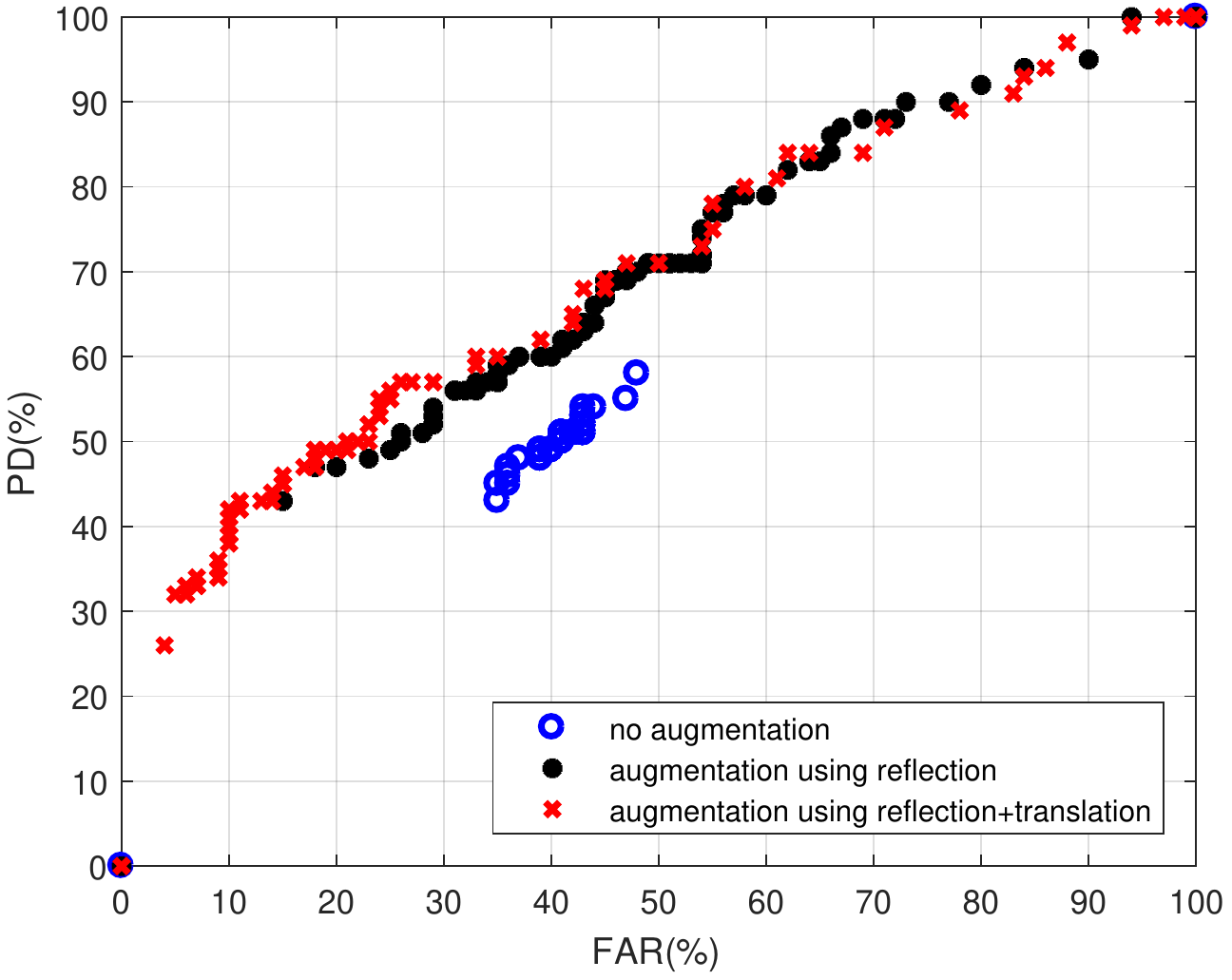}
\label{fig:augmentation-25}}
\subfloat[][$-26$dB]{
\includegraphics[width=0.15\textwidth]{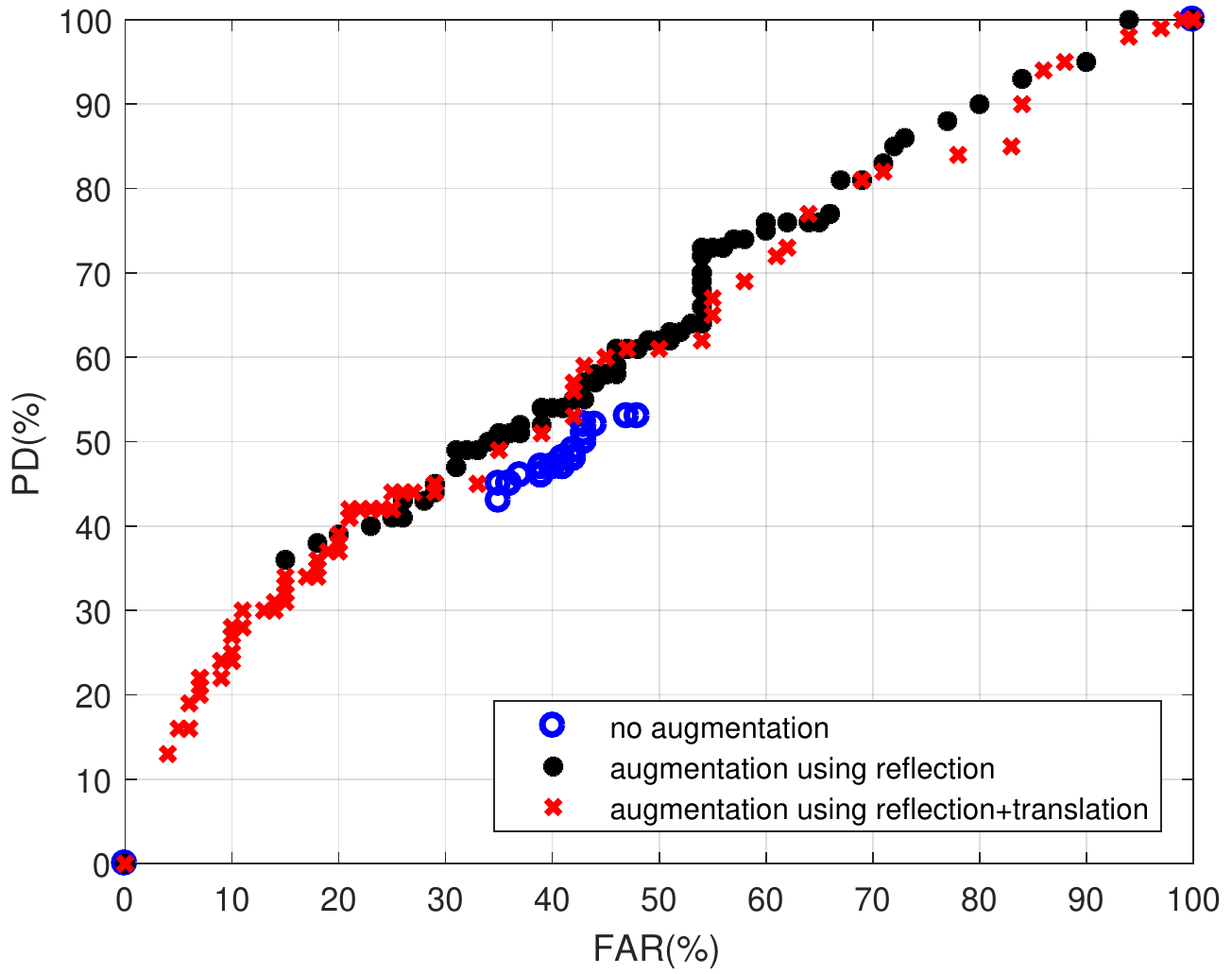}
\label{fig:augmentation-26}}
\caption{(a)Lofargram examples: noise only $H_0$ lofargram and $H_1$ lofargrams under different SNRs. (b)-(f) ROC curves for data augmentation: using vertical and horizontal reflections and translation with $4$ random $224\times224$ patches from the $500\times256$ lofargrams in different SNRs.}
\label{fig:augmentation}
\end{figure}

Deep neural networks based tasks should be improved with training data augmentation. We first generate vertical and horizontal reflections for each of the training lofargram. This increases the size of our training set by a factor of $8$. Then we further augment the training set by translation that is extracting $4$ random $224\times224$ patches from the $500\times256$ lofargrams, and resulting in an augmented dataset $32$ times larger than the original one. Fig.~\ref{fig:augmentation-22}-\ref{fig:augmentation-26} show the precision-recall curves using AlexNet, and data augmentation consistently improves the performances in varied SNRs.

%We also try to augment the training set with varied SNR settings, and experimental results show that such an operation decrease our performance by more than 0.5 points. This may be because .
%\begin{figure}
%\centering
%\subfloat[][deconv]{
%\includegraphics[width=0.15\textwidth]{vis981_bp}}
%\subfloat[][Line Location Accuracy]{
%\includegraphics[width=0.15\textwidth]{vis981_guidedbp}}
%\caption{The two .}
%\label{fig:evametric}
%\end{figure}

\subsubsection{ConvNet Depth}
\begin{figure}
\centering
\subfloat[][$-22$dB]{
\includegraphics[width=0.15\textwidth]{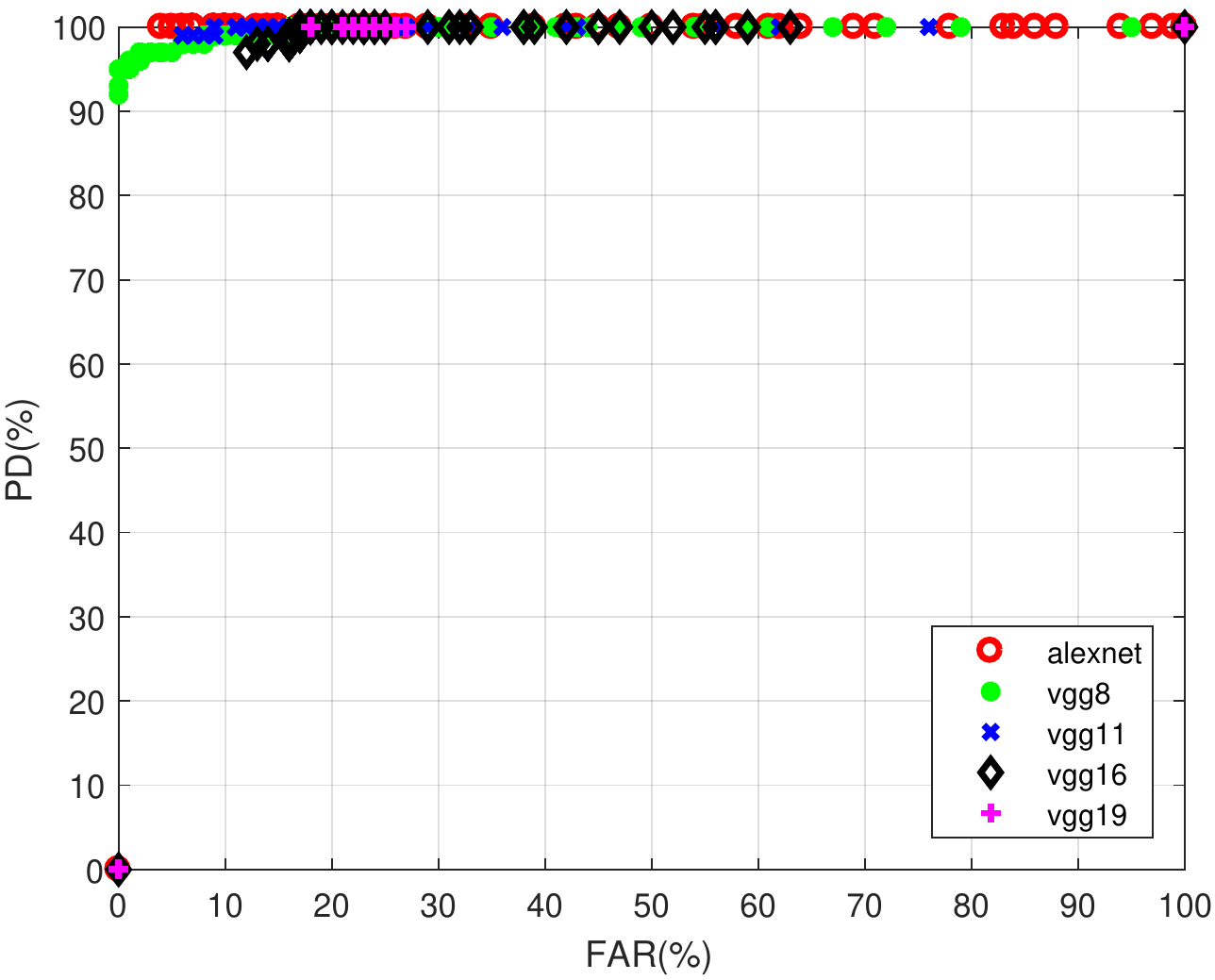}}
\subfloat[][$-23$dB]{
\includegraphics[width=0.15\textwidth]{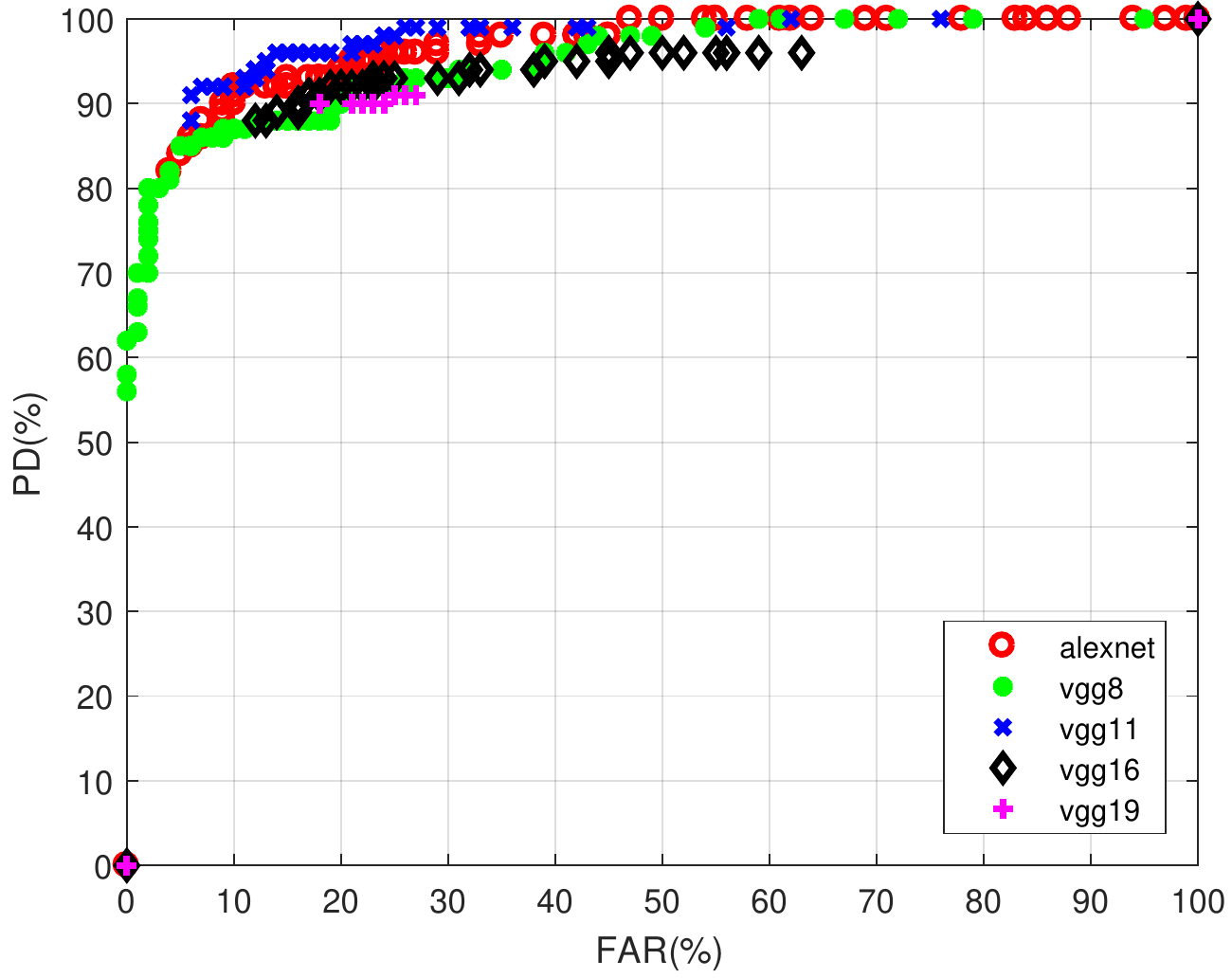}}
\subfloat[][$-24$dB]{
\includegraphics[width=0.15\textwidth]{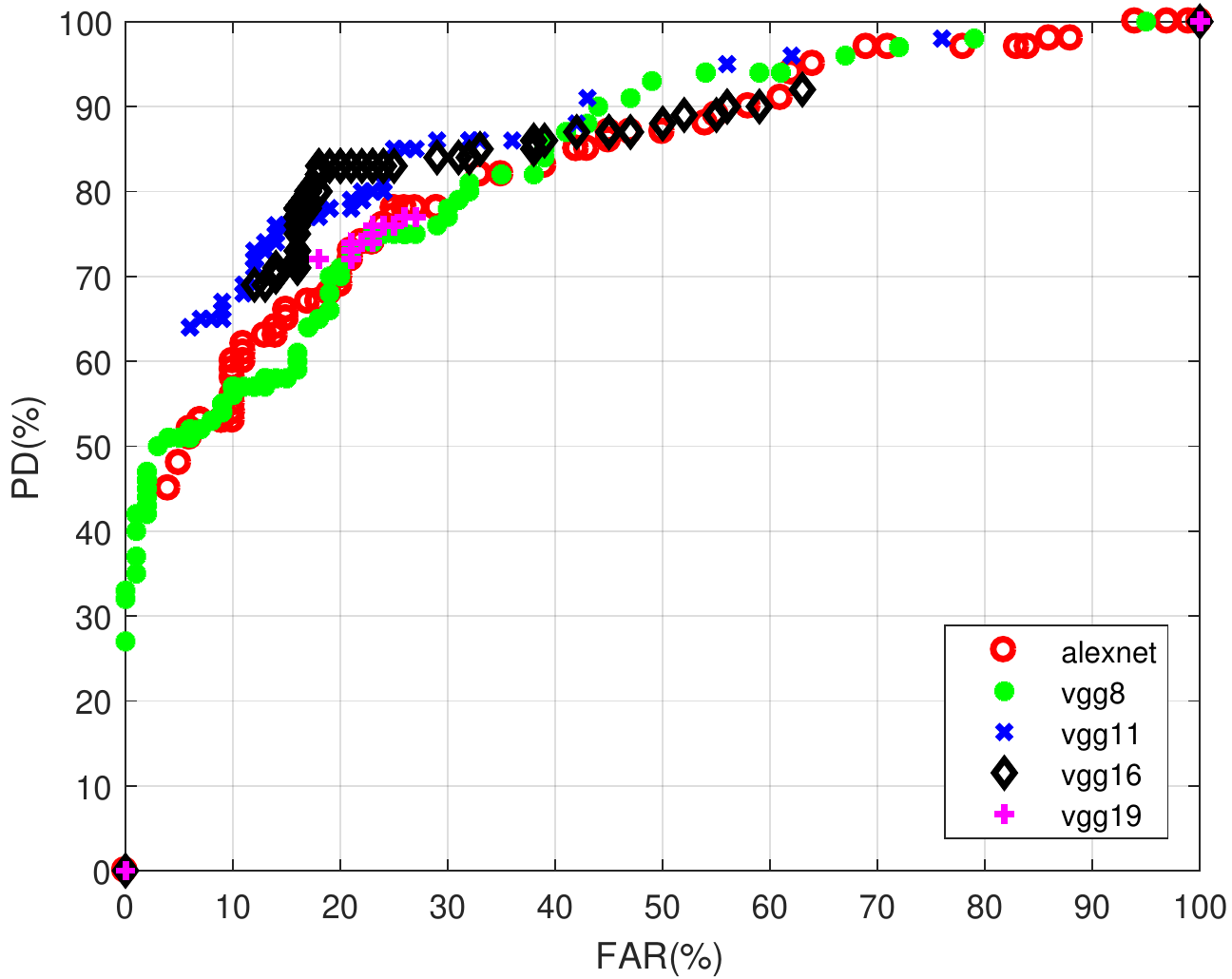}}\\
\subfloat[][$-25$dB]{
\includegraphics[width=0.15\textwidth]{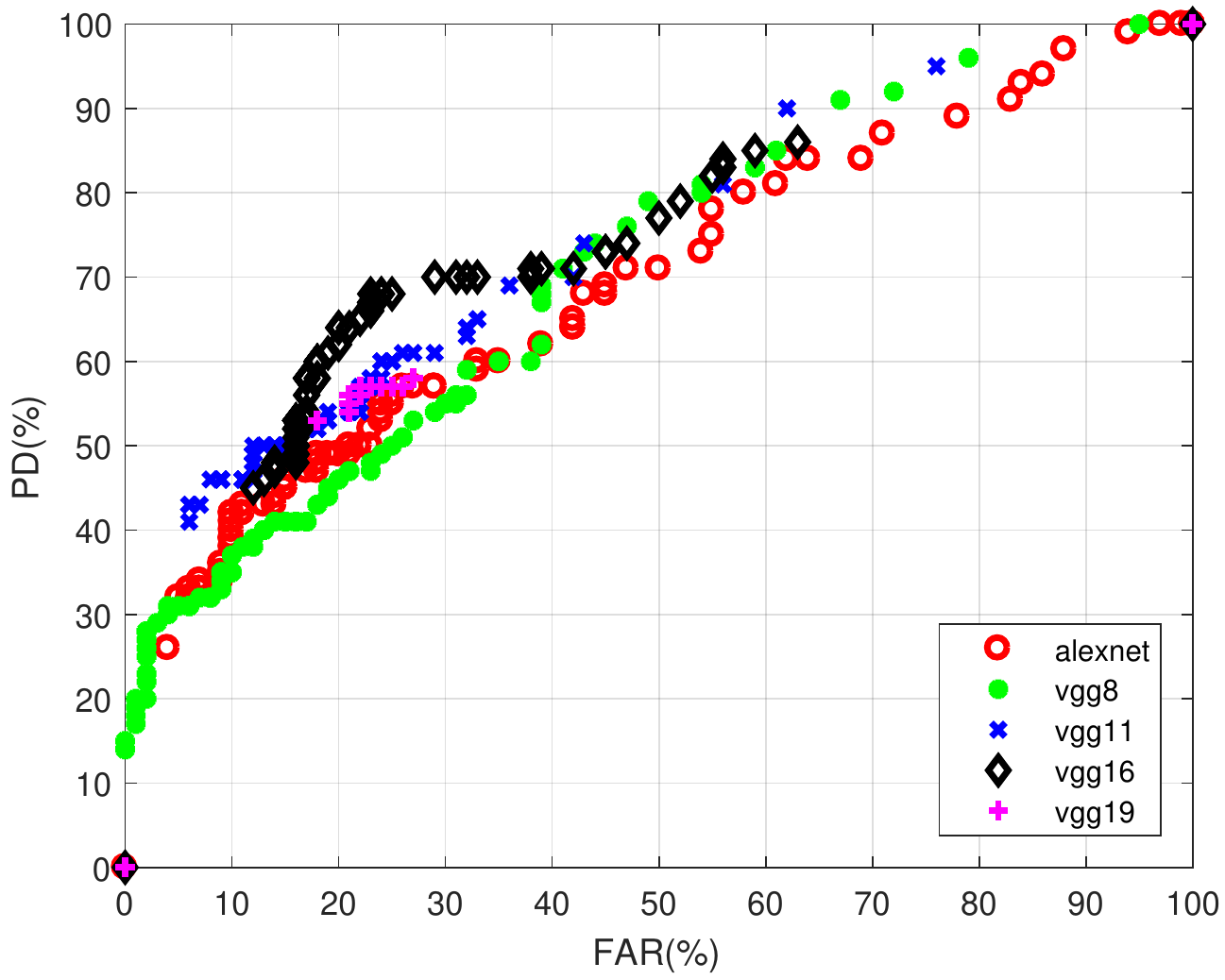}}
\subfloat[][$-26$dB]{
\includegraphics[width=0.15\textwidth]{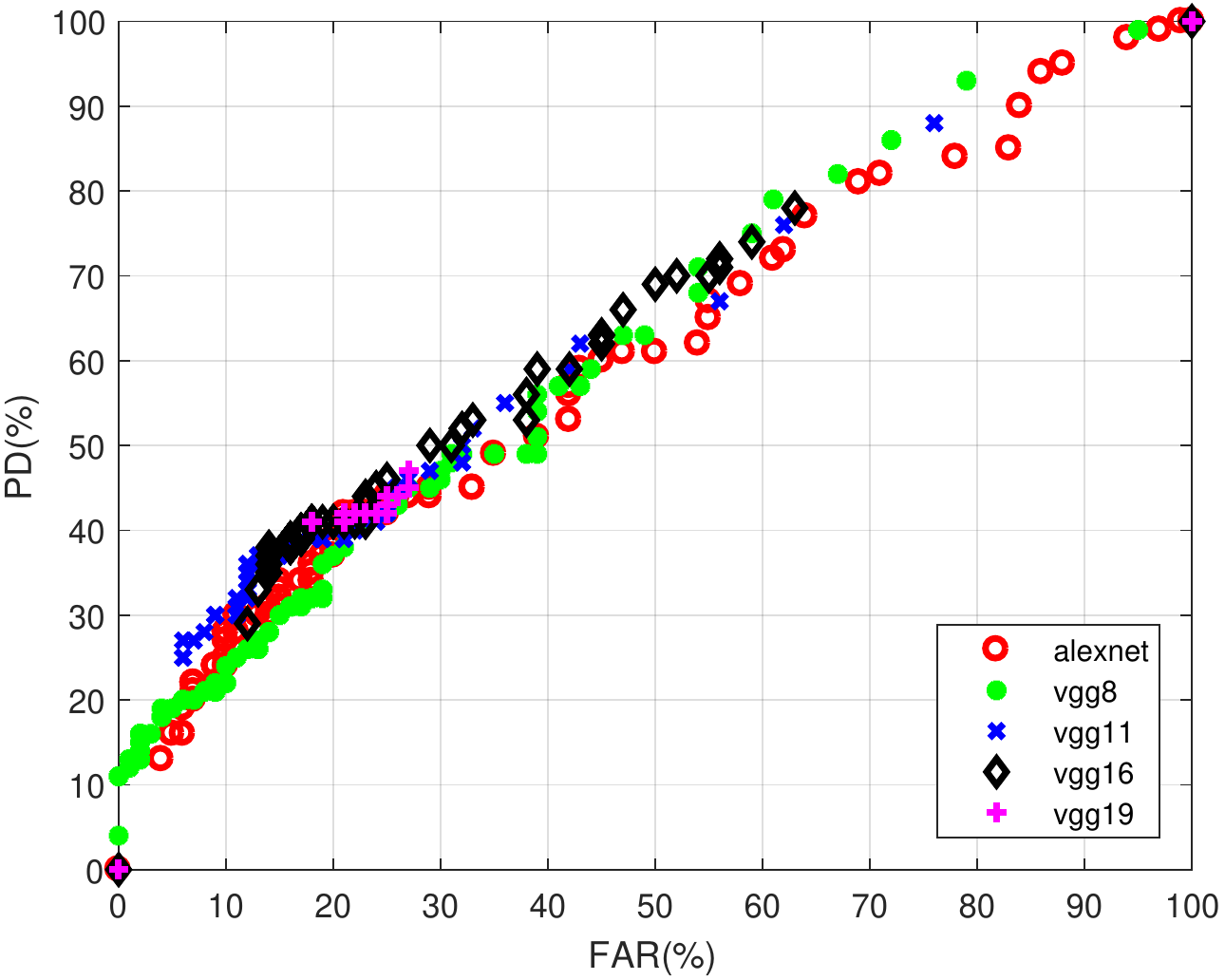}}
\subfloat[][LLA]{
\includegraphics[width=0.15\textwidth]{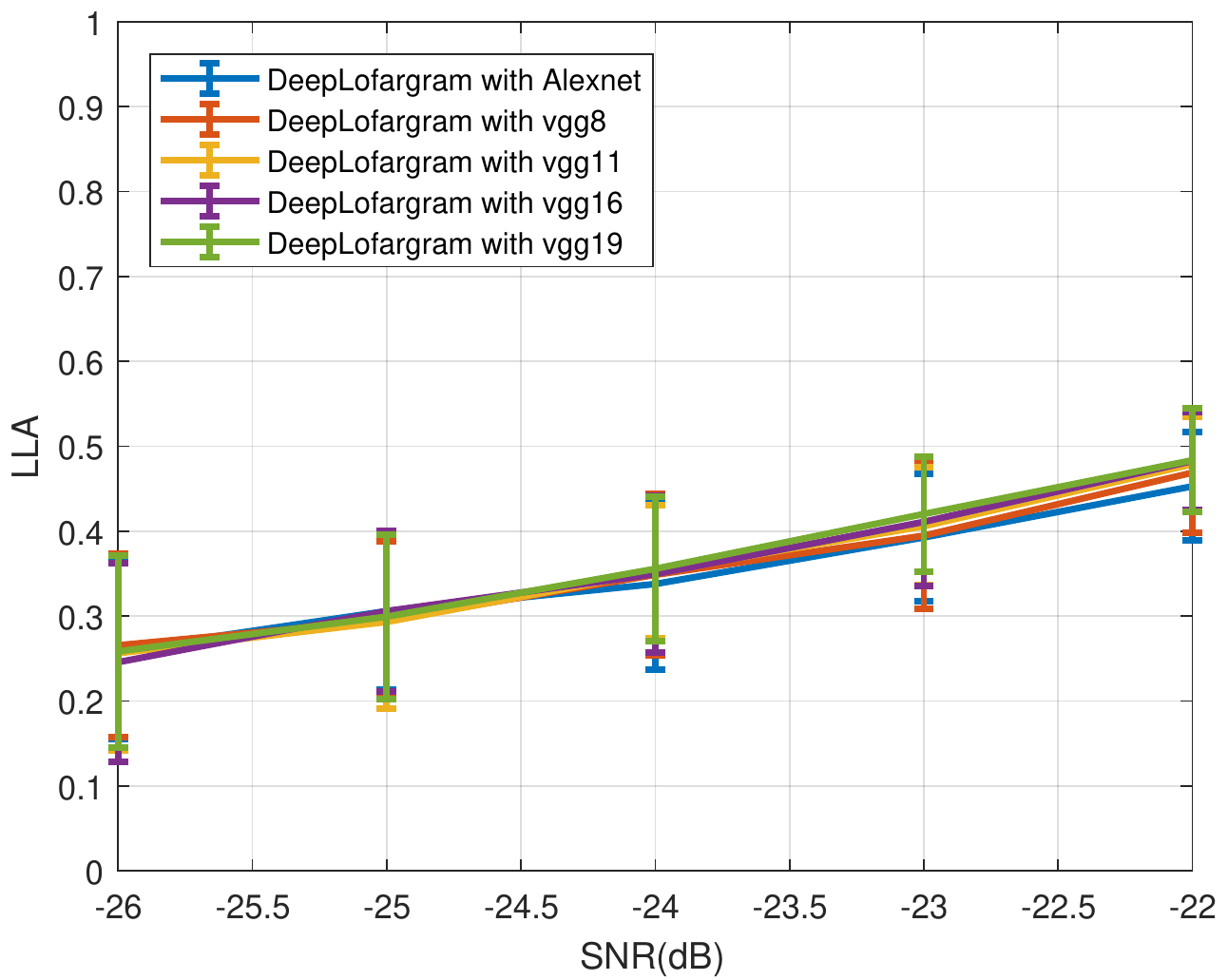}
\label{fig:depth-lla}}
\caption{ConvNet depths: (a)-(e) ROC curves and (f) line location accuracies in different SNRs.}
\label{fig:depth}
\end{figure}

To examine the effect of increasing convolutional network depth, we also replace Alexnet ($8$ layers) with VggNet$X$~\cite{SimonyanZ2014a} as the backbone of our DeepLofargram, where $X\in\{8,11,16,19\}$ represents the number of layers of VggNet. Fig.~\ref{fig:depth} compares the precision-recall curves and line location accuracies.  Convolutional network with different depth show comparable performances in dim frequency line detection and recovery.

\subsubsection{Visualization Technologies}
\begin{figure}
\centering
\subfloat[][Alexnet]{
\includegraphics[width=0.15\textwidth]{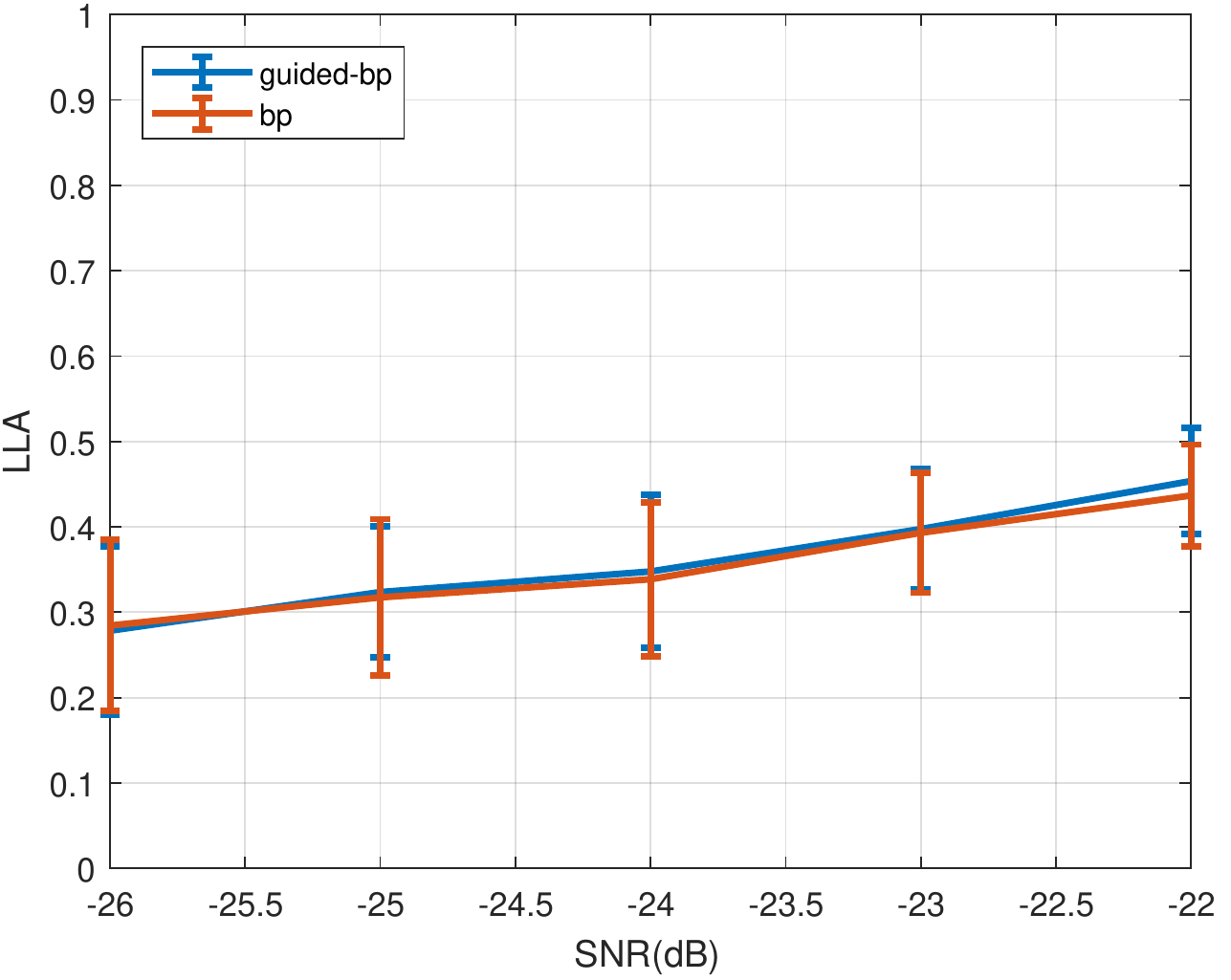}
\label{fig:visAlex}}
\subfloat[][VggNet$8$]{
\includegraphics[width=0.15\textwidth]{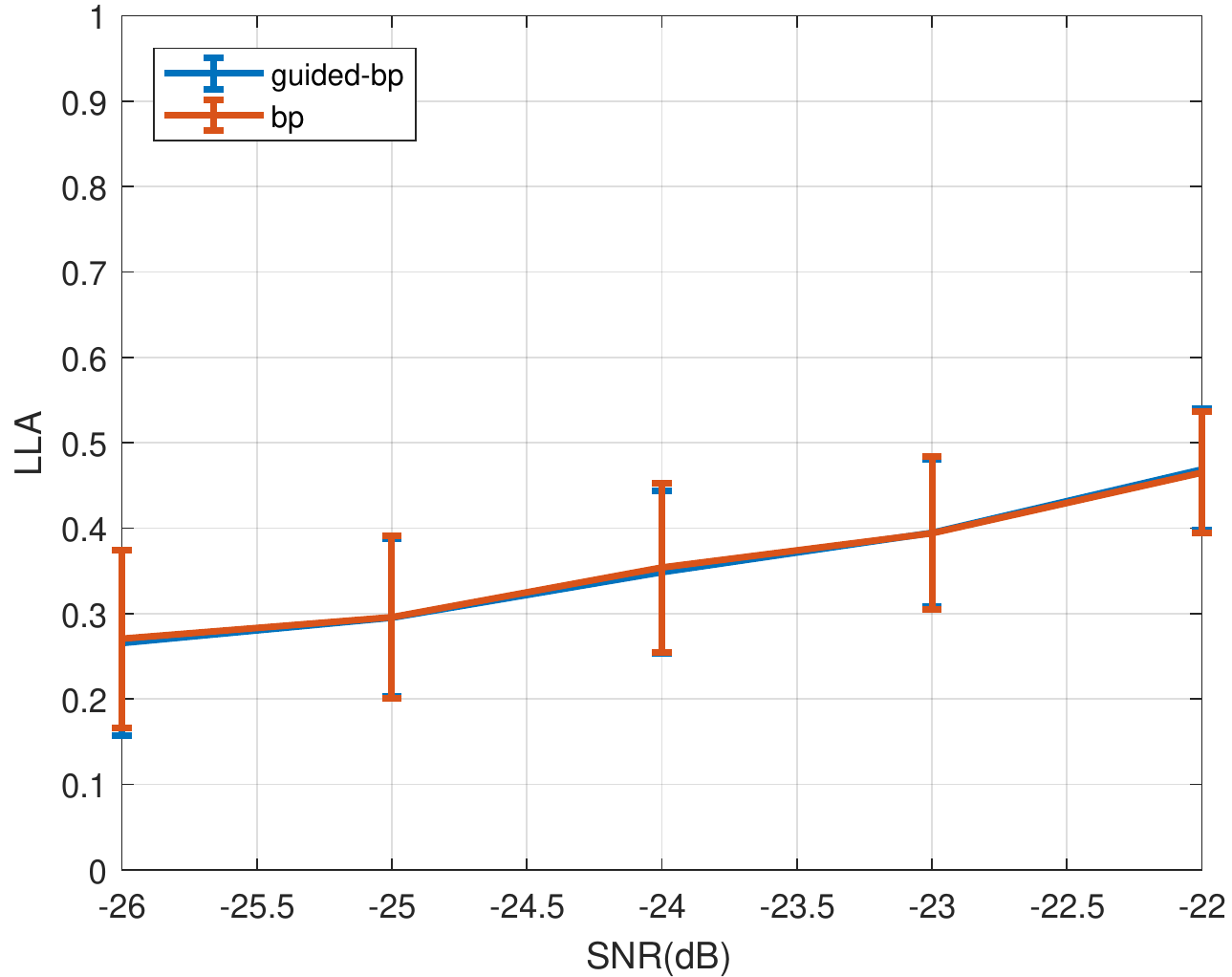}
\label{fig:visVgg8}}
\subfloat[][VggNet$11$]{
\includegraphics[width=0.15\textwidth]{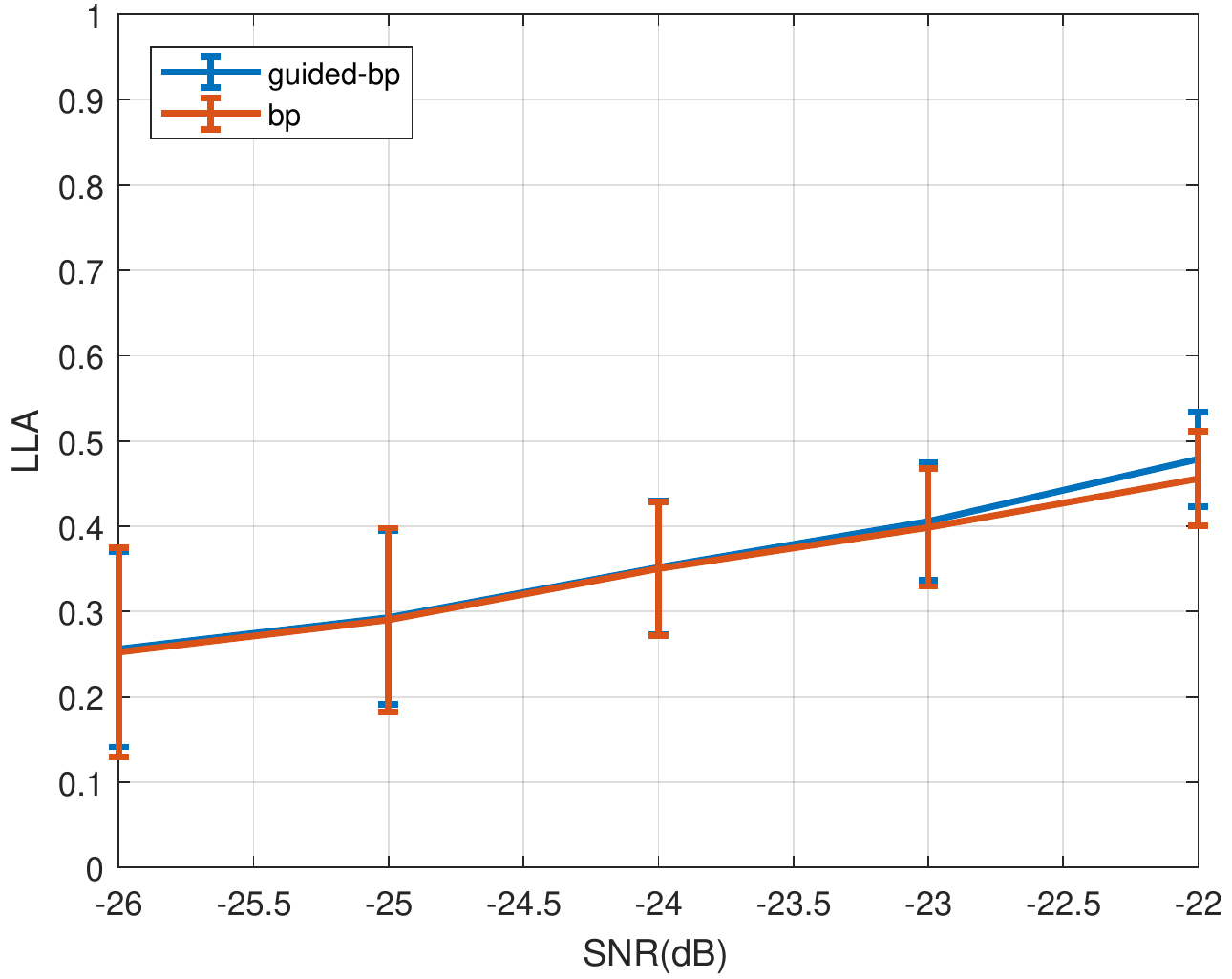}
\label{fig:visVgg11}}\\
\subfloat[][VggNet$16$]{
\includegraphics[width=0.15\textwidth]{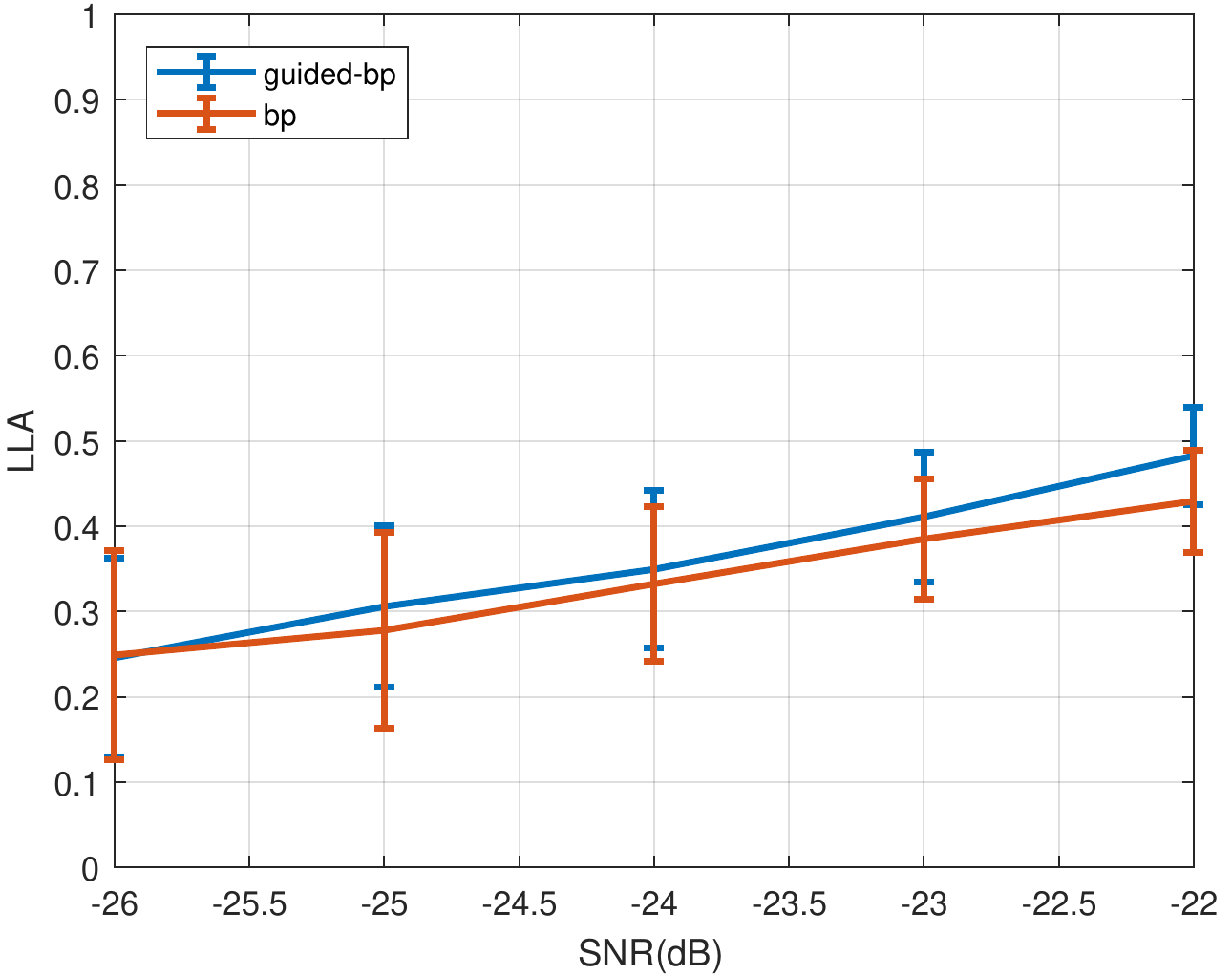}
\label{fig:visVgg16}}
\subfloat[][VggNet$19$]{
\includegraphics[width=0.15\textwidth]{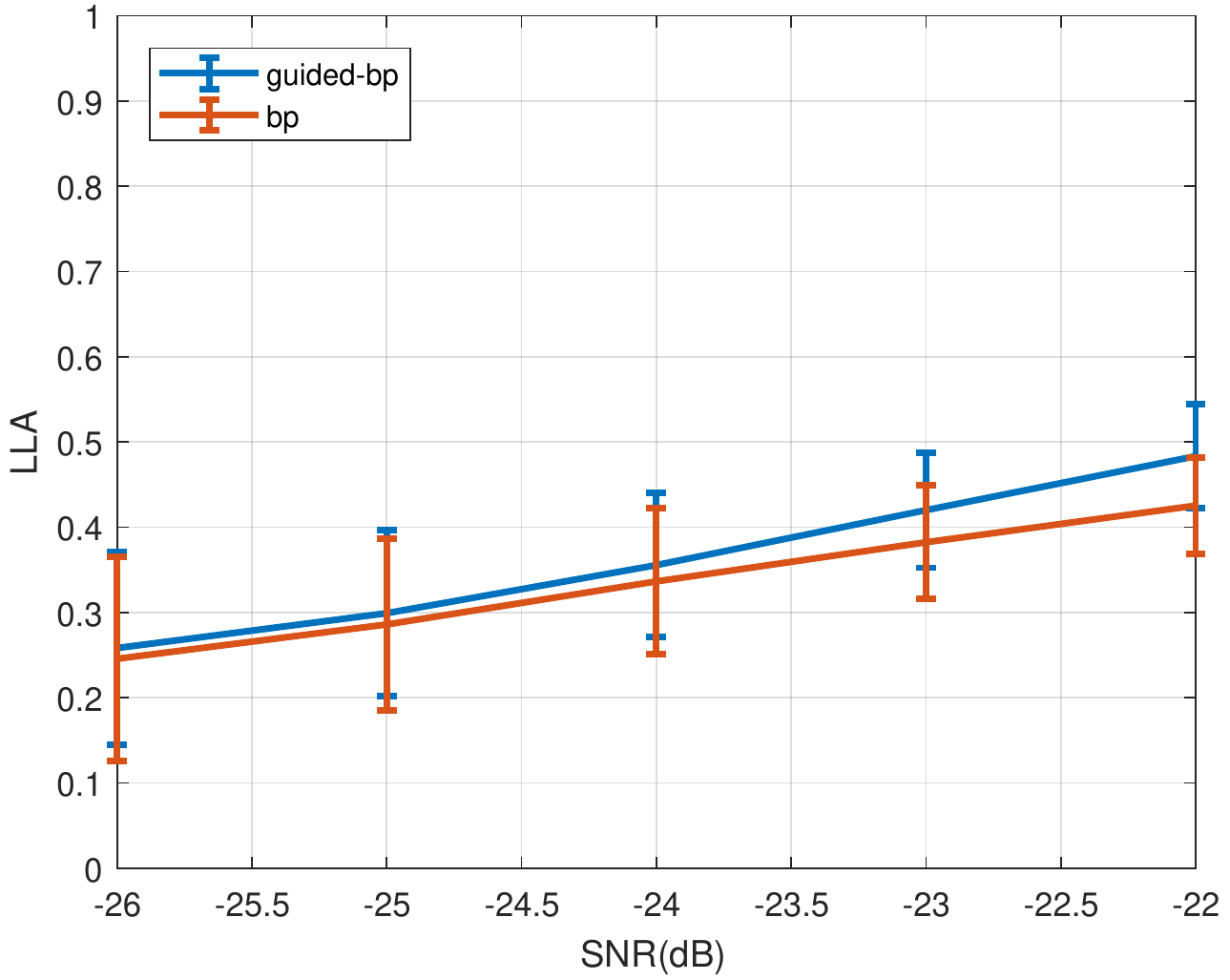}
\label{fig:visVgg19}}
\subfloat[][The state-of-the-art]{
\includegraphics[width=0.15\textwidth]{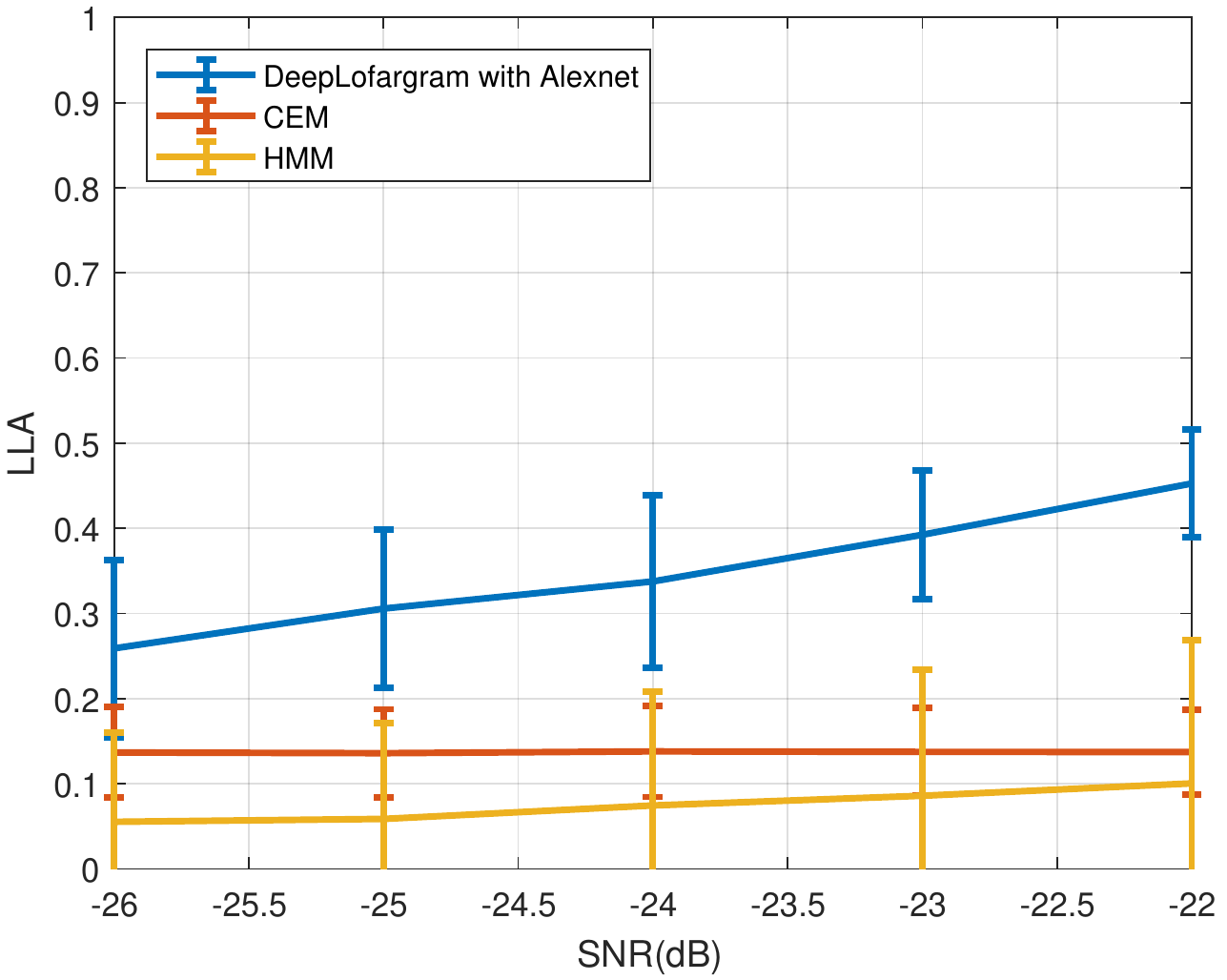}
\label{fig:lla}}
\caption{Comparison of LLA for (a)-(e) visualization technologies (guided backpropagation versus backpropagation) and (f) the state-of-the-art methods in different SNRs.}
\label{fig:vis}
\end{figure}

With the same backbones, we run two visualization technologies, backpropagation (BP)~\cite{SimonyanVZ2014} and guided backpropagation~\cite{SpringenbergDBR2015}, to explore the best output. The line location accuracies of each methods in varied SNR has been shown in Fig.~\ref{fig:visAlex}-\ref{fig:visVgg19}. Guided BP with deeper network leads slightly better LLA than BP. These improvements become more obvious in higher SNR.

\subsection{Comparison with the State-of-the-Art}
\begin{figure}
\centering
\subfloat[][Alex-22]{
\includegraphics[width=0.09\textwidth]{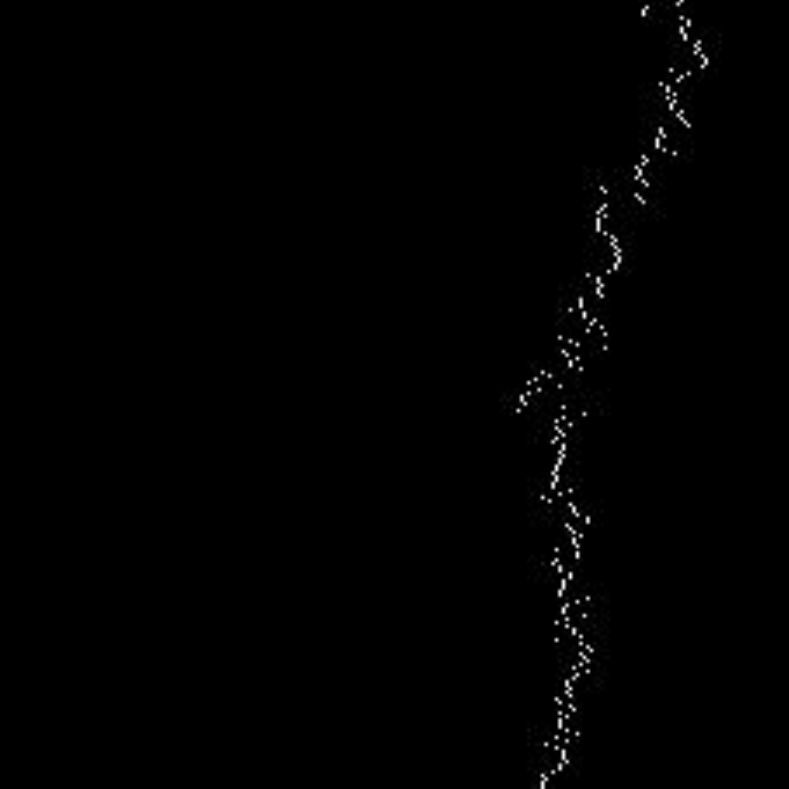}}
\subfloat[][Alex-23]{
\includegraphics[width=0.09\textwidth]{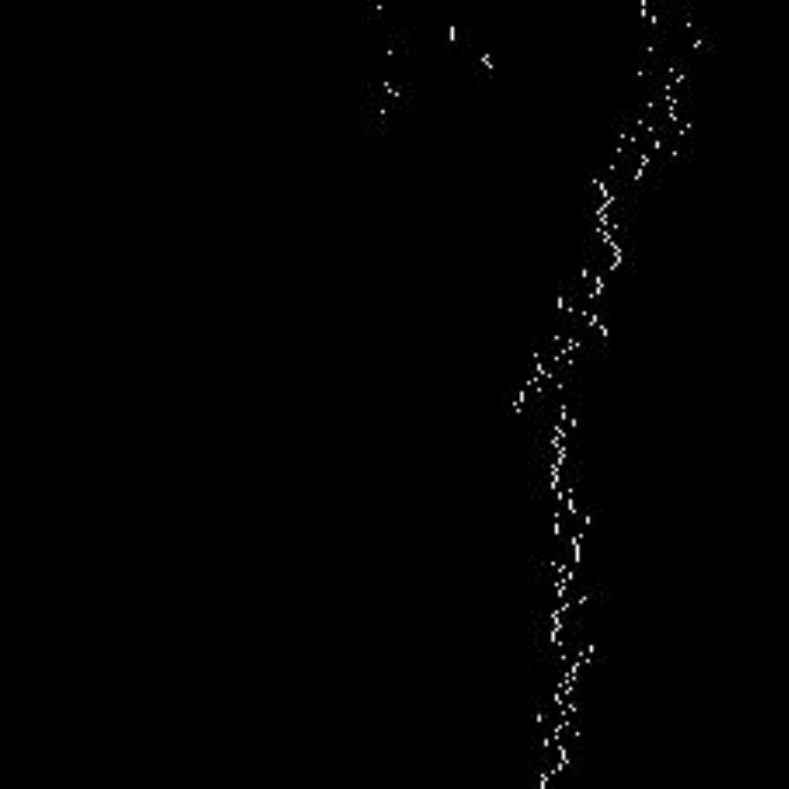}}
\subfloat[][Alex-24]{
\includegraphics[width=0.09\textwidth]{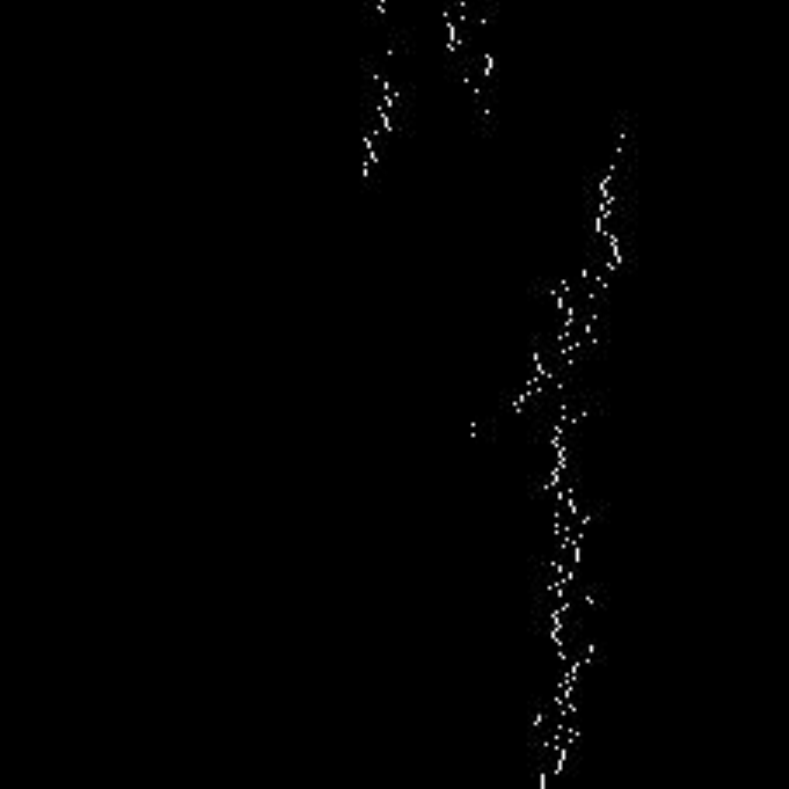}}
\subfloat[][Alex-25]{
\includegraphics[width=0.09\textwidth]{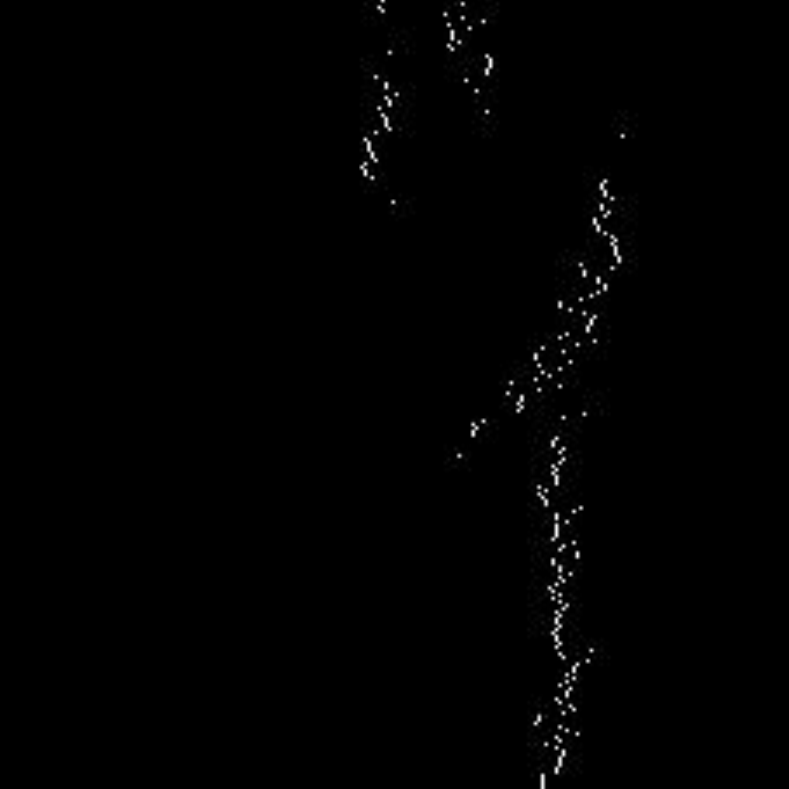}}
\subfloat[][Alex-26]{
\includegraphics[width=0.09\textwidth]{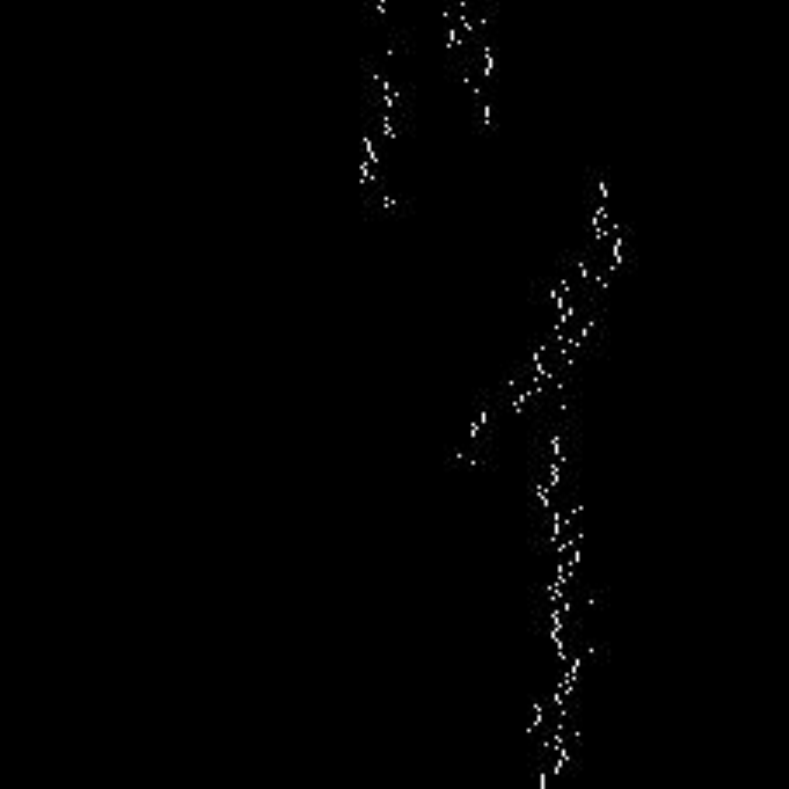}}

\subfloat[][Vgg8-22]{
\includegraphics[width=0.09\textwidth]{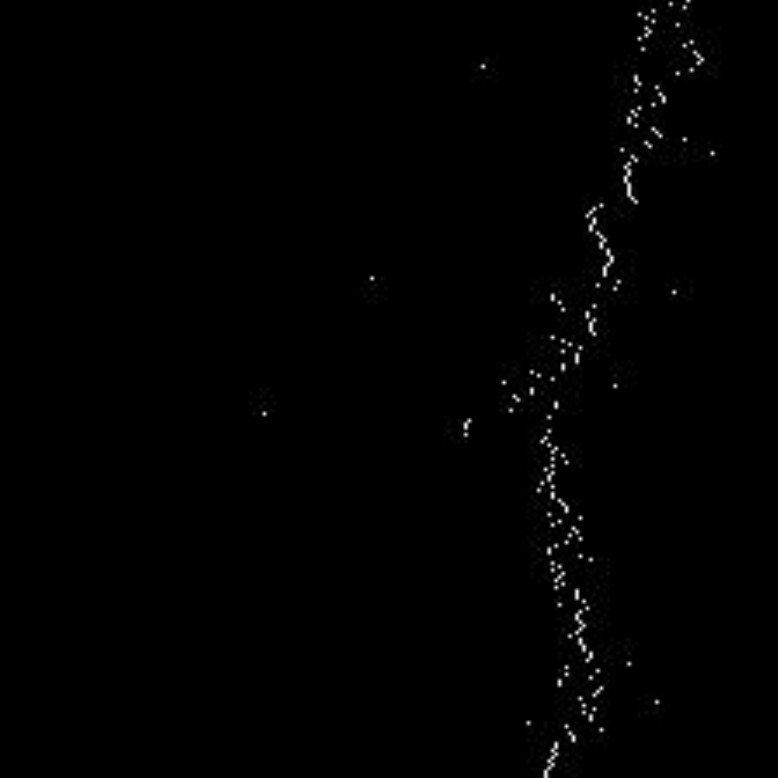}}
\subfloat[][Vgg8-23]{
\includegraphics[width=0.09\textwidth]{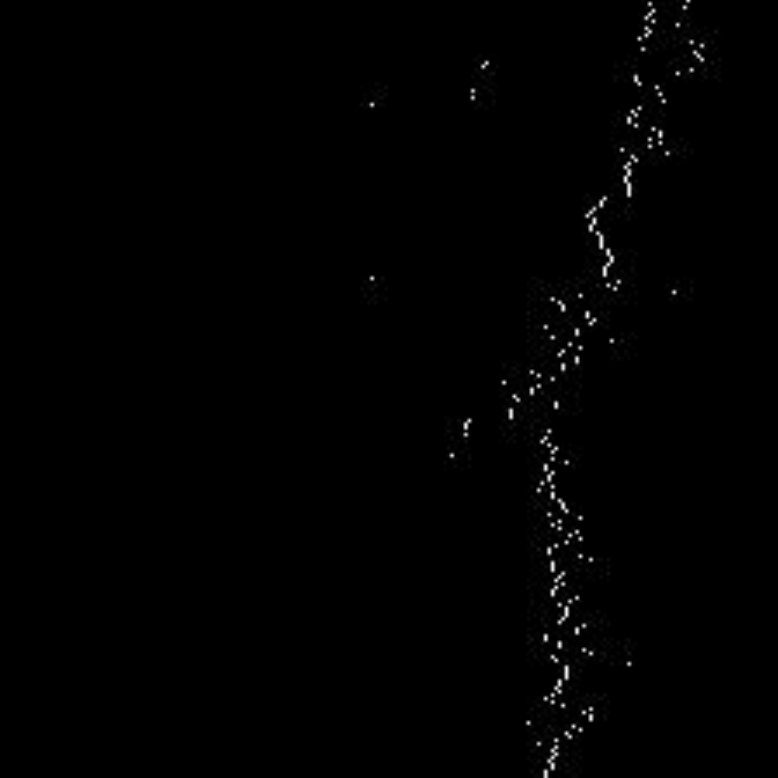}}
\subfloat[][Vgg8-24]{
\includegraphics[width=0.09\textwidth]{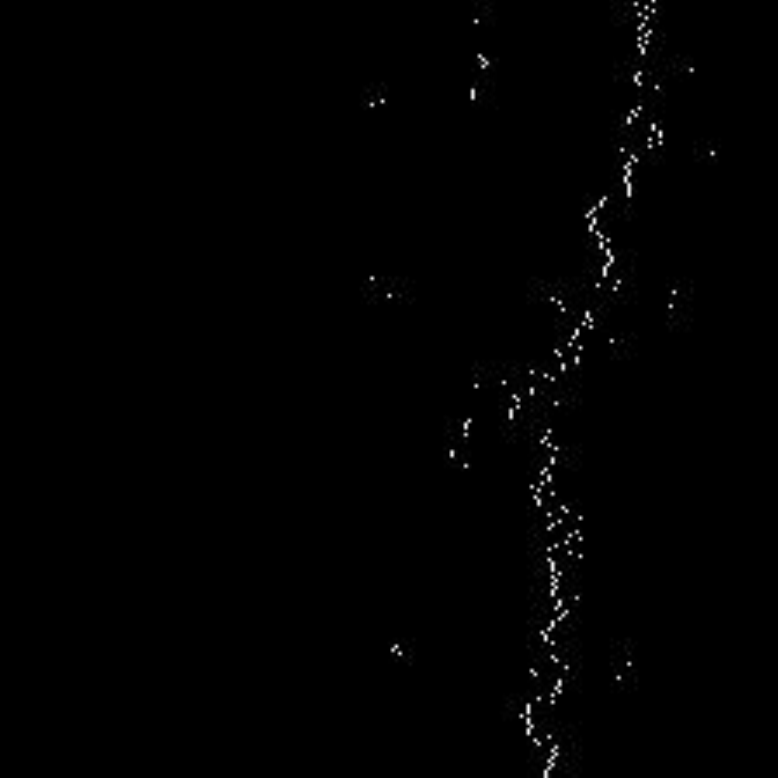}}
\subfloat[][Vgg8-25]{
\includegraphics[width=0.09\textwidth]{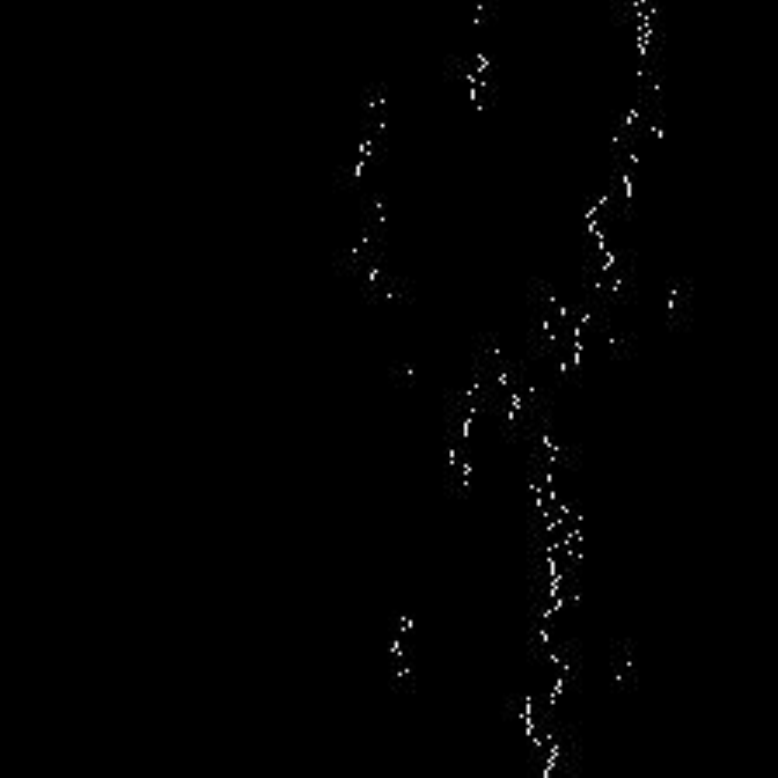}}
\subfloat[][Vgg8-26]{
\includegraphics[width=0.09\textwidth]{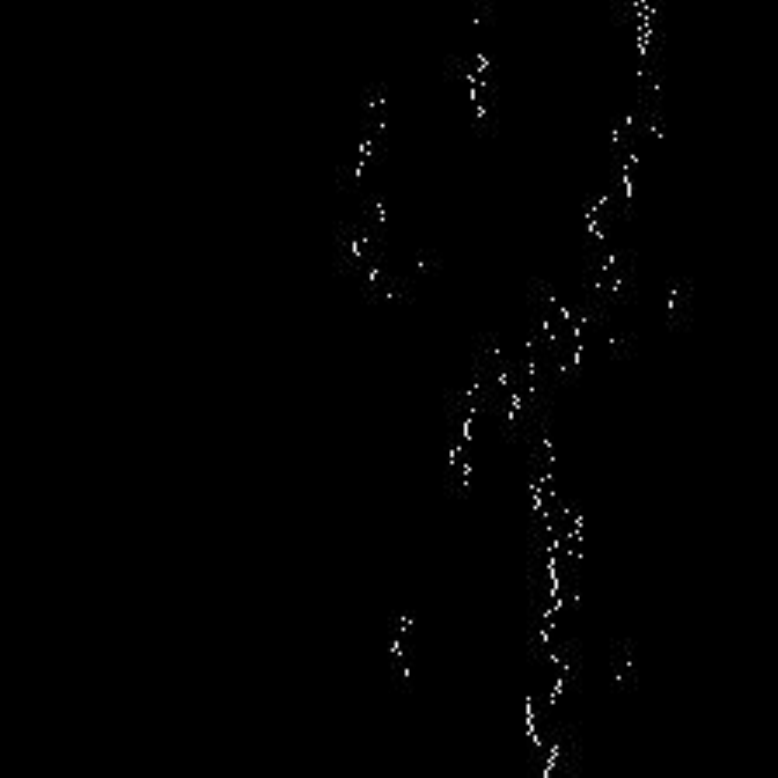}}

\subfloat[][Vgg11-22]{
\includegraphics[width=0.09\textwidth]{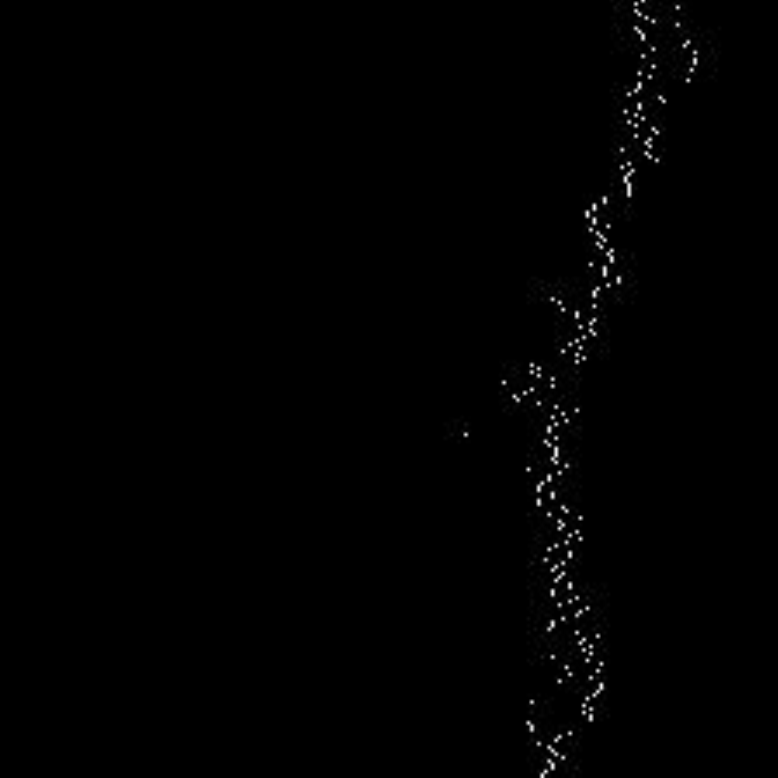}}
\subfloat[][Vgg11-23]{
\includegraphics[width=0.09\textwidth]{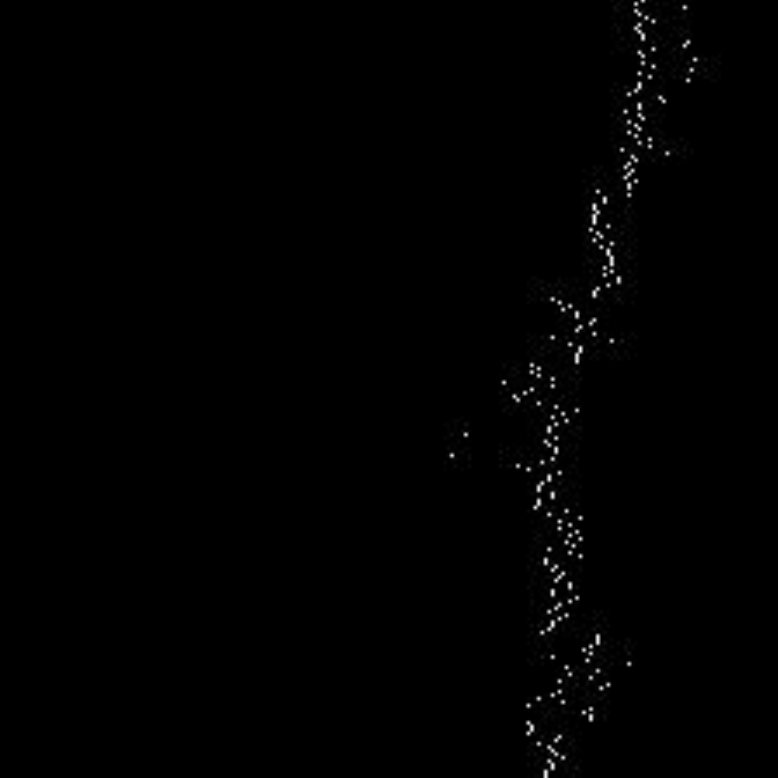}}
\subfloat[][Vgg11-24]{
\includegraphics[width=0.09\textwidth]{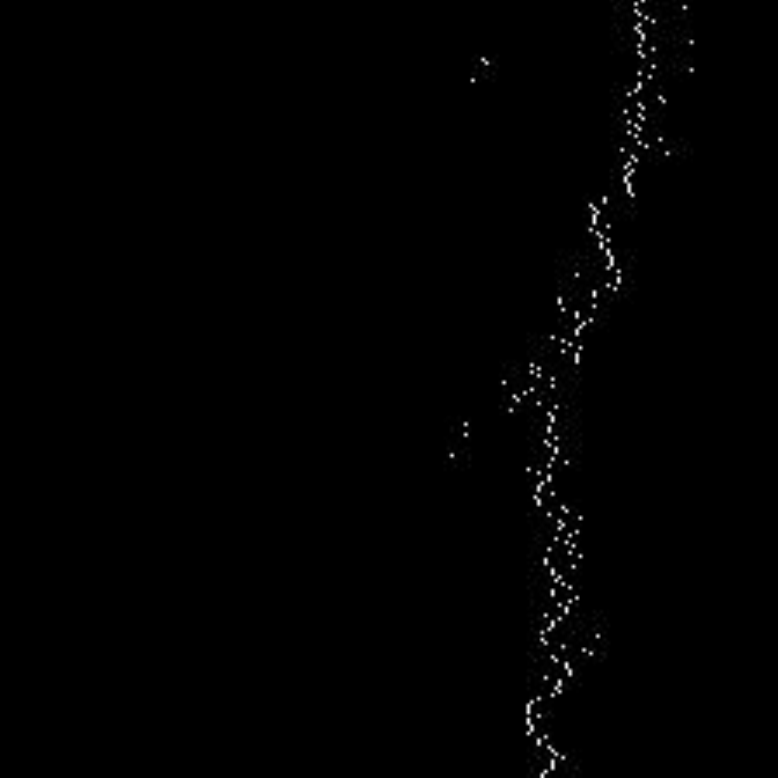}}
\subfloat[][Vgg11-25]{
\includegraphics[width=0.09\textwidth]{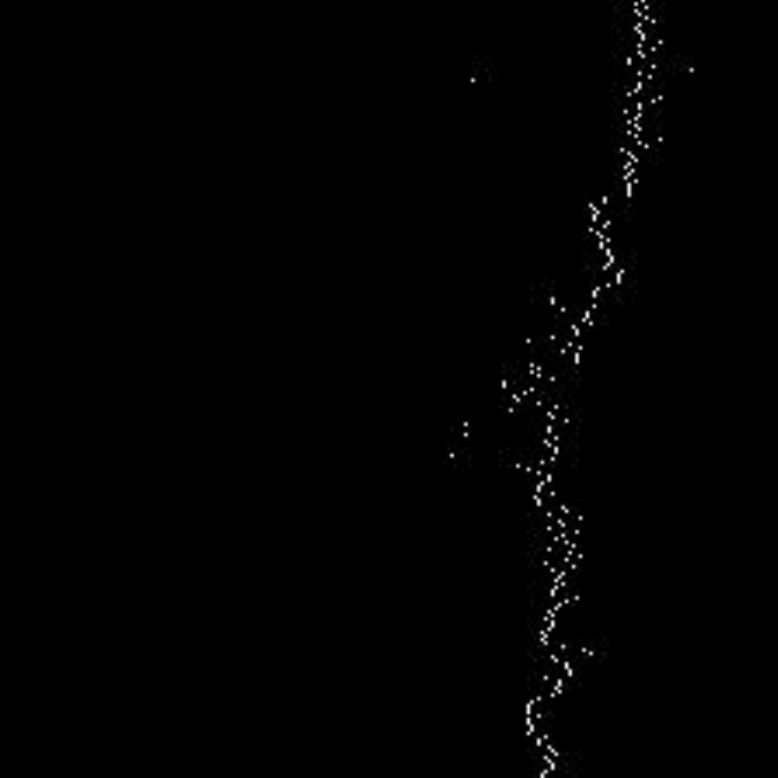}}
\subfloat[][Vgg11-26]{
\includegraphics[width=0.09\textwidth]{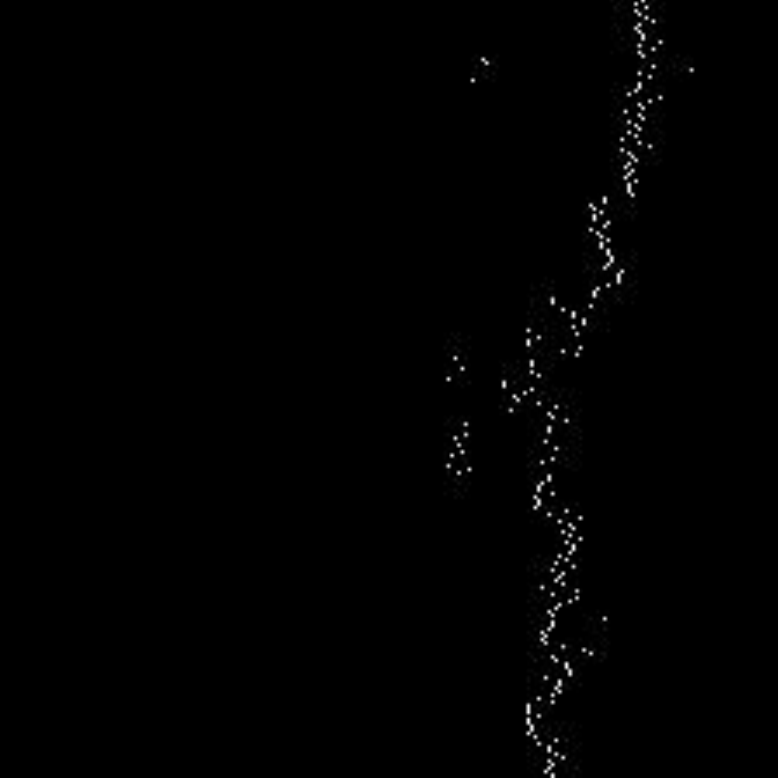}}

\subfloat[][Vgg16-22]{
\includegraphics[width=0.09\textwidth]{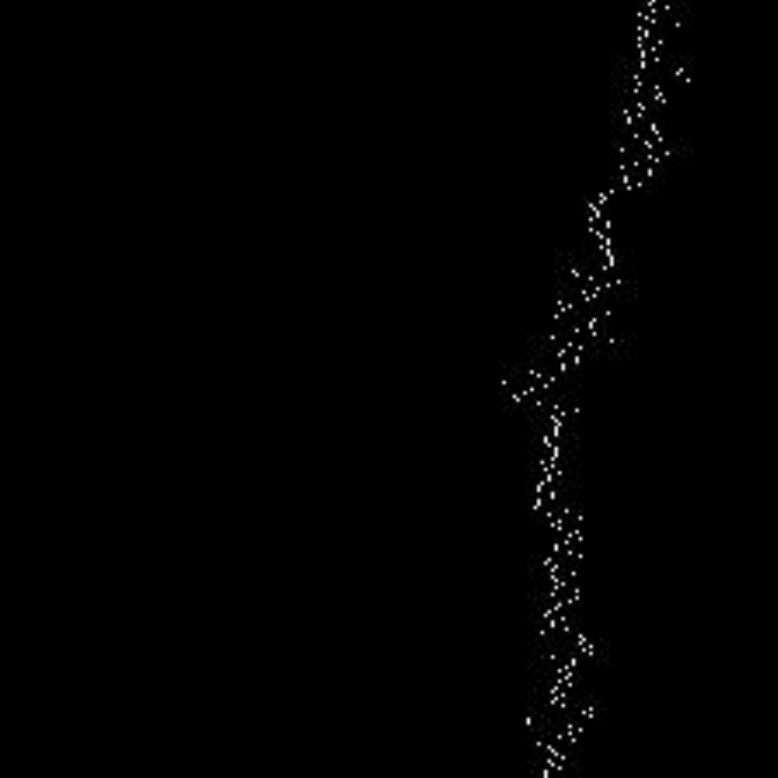}}
\subfloat[][Vgg16-23]{
\includegraphics[width=0.09\textwidth]{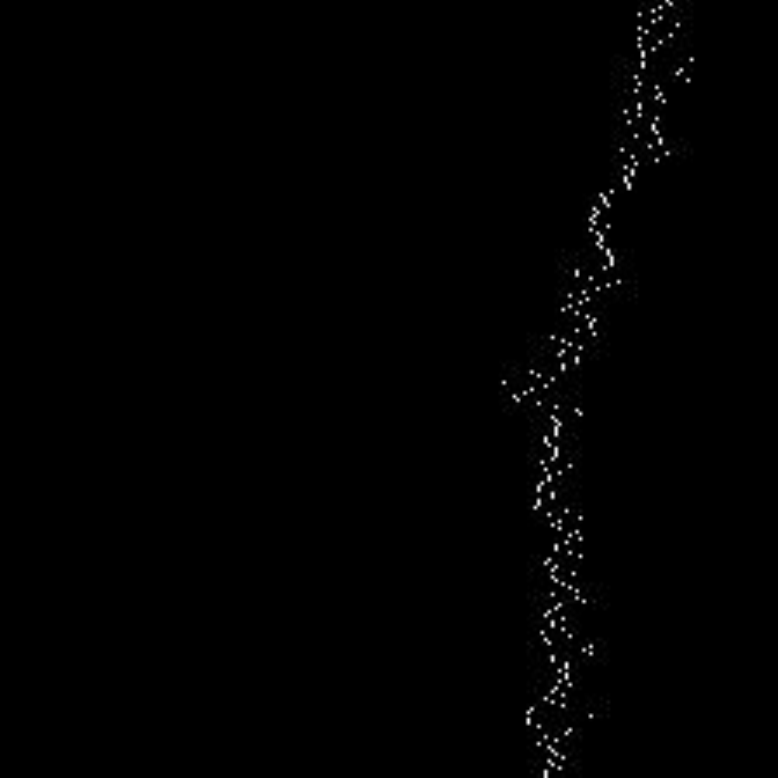}}
\subfloat[][Vgg16-24]{
\includegraphics[width=0.09\textwidth]{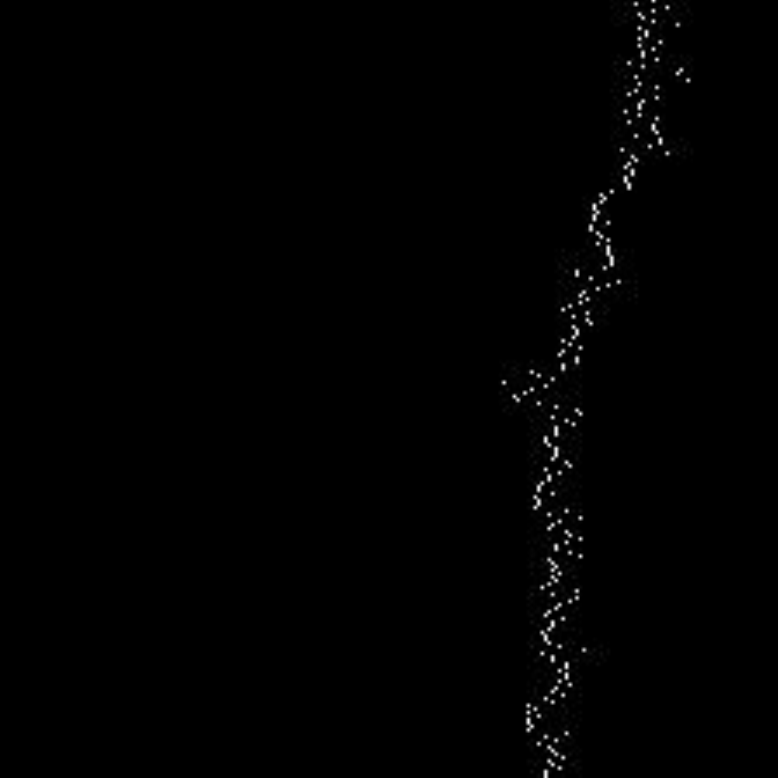}}
\subfloat[][Vgg16-25]{
\includegraphics[width=0.09\textwidth]{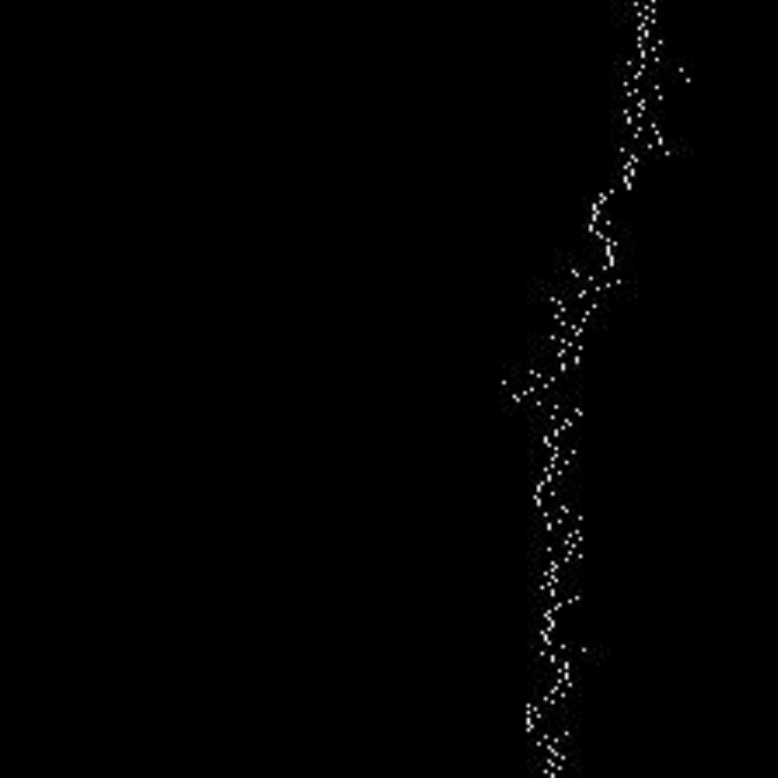}}
\subfloat[][Vgg16-26]{
\includegraphics[width=0.09\textwidth]{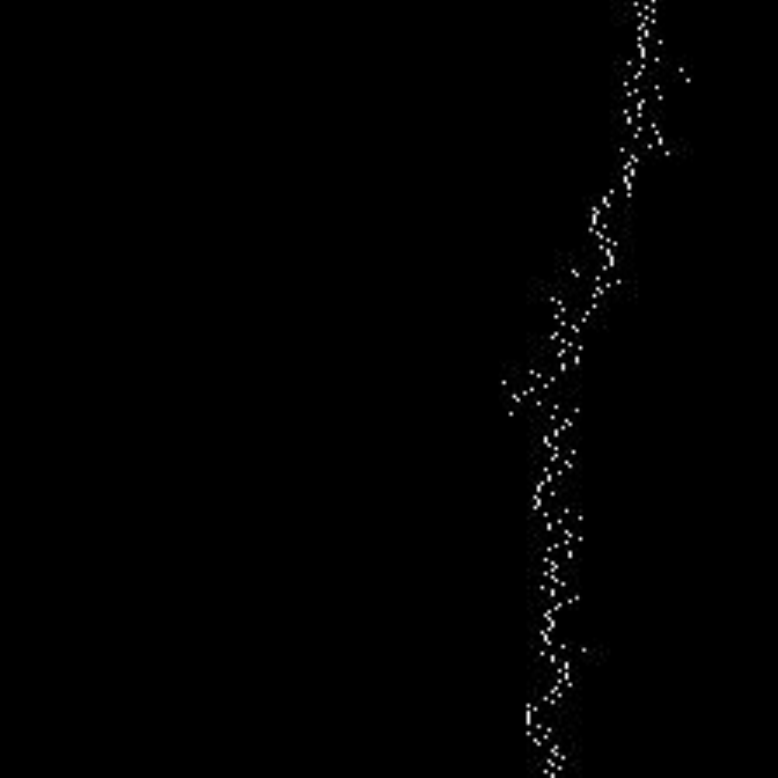}}

\subfloat[][Vgg19-22]{
\includegraphics[width=0.09\textwidth]{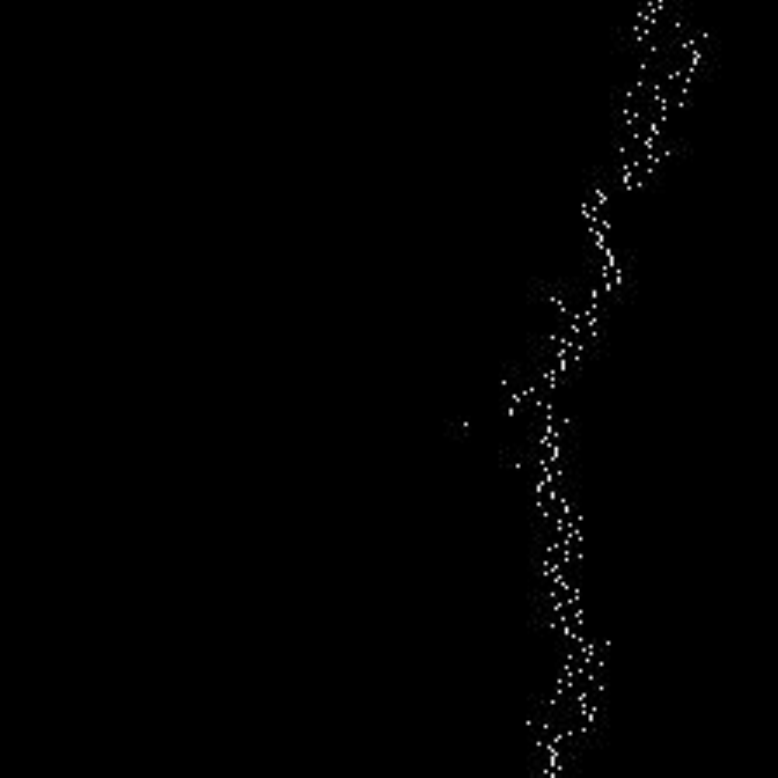}}
\subfloat[][Vgg19-23]{
\includegraphics[width=0.09\textwidth]{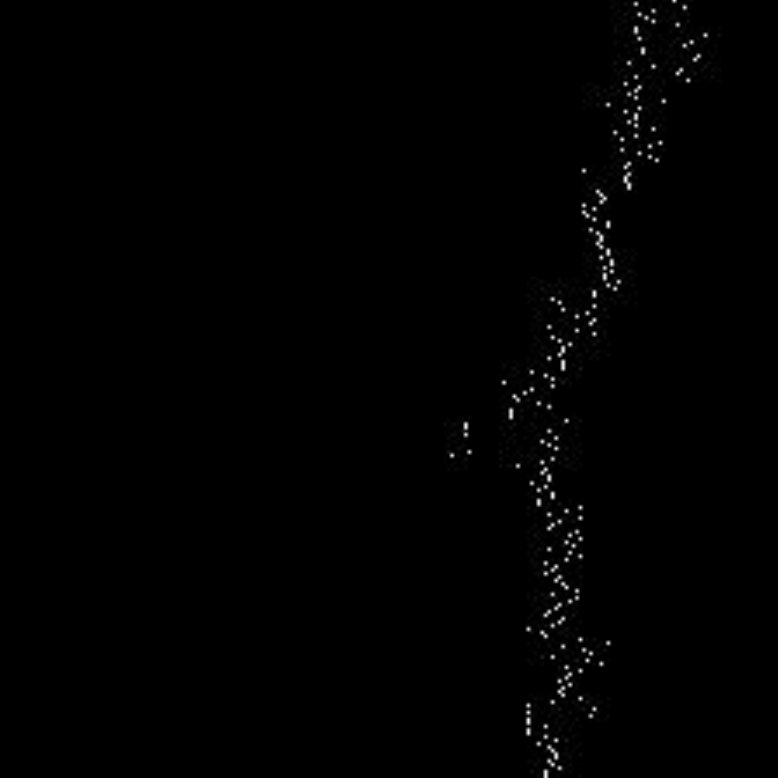}}
\subfloat[][Vgg19-24]{
\includegraphics[width=0.09\textwidth]{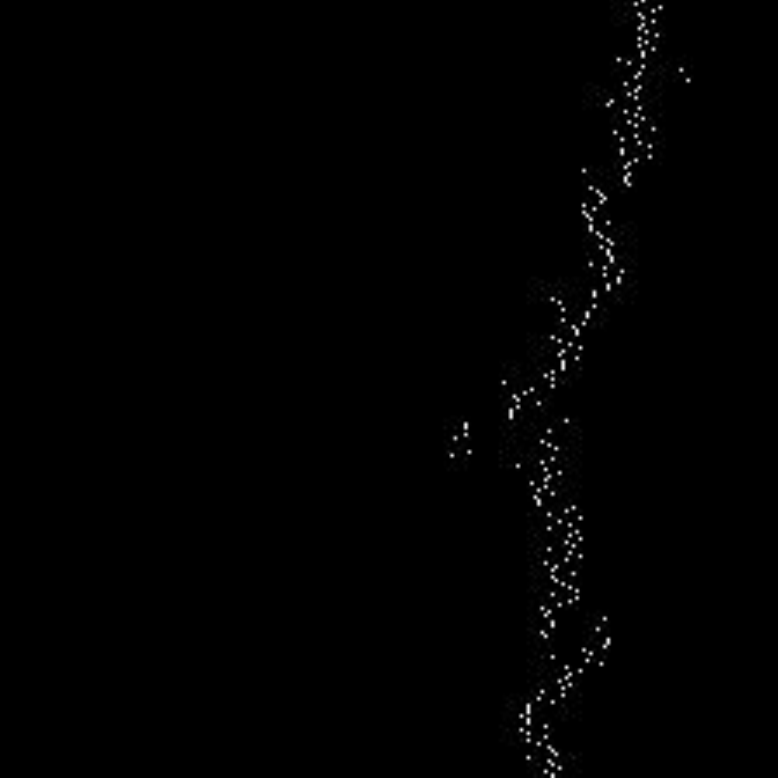}}
\subfloat[][Vgg19-25]{
\includegraphics[width=0.09\textwidth]{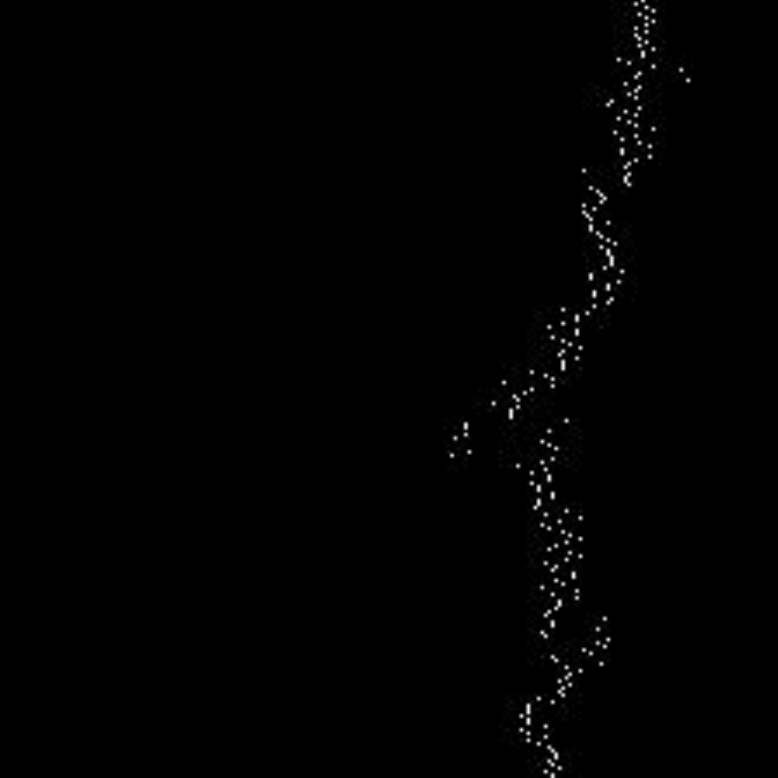}}
\subfloat[][Vgg19-26]{
\includegraphics[width=0.09\textwidth]{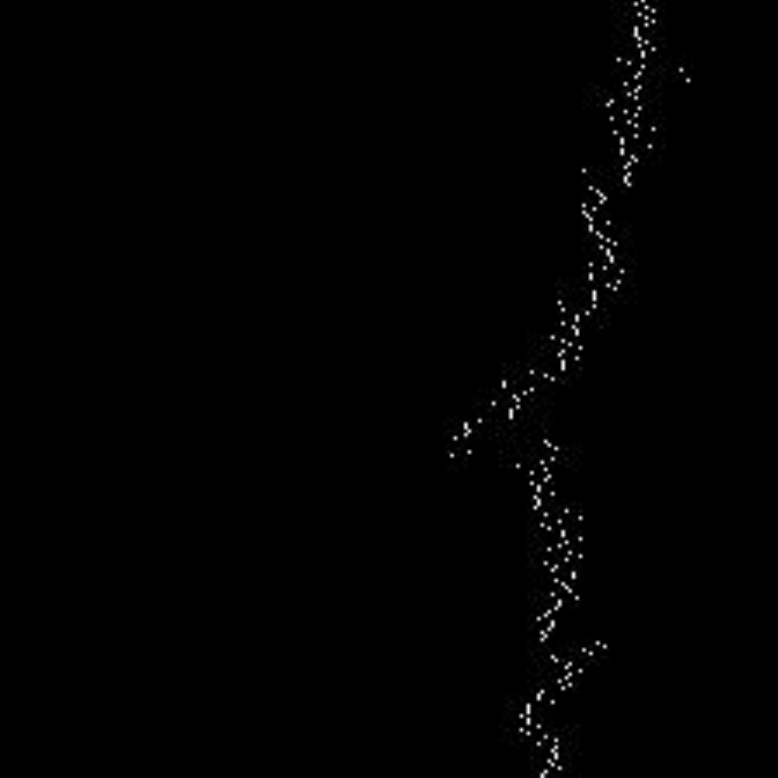}}

%\subfloat[CEM-22]{
%\includegraphics[width=0.09\textwidth]{HMM-22db}}
%\subfloat[][HMM-23]{
%\includegraphics[width=0.09\textwidth]{HMM-23db}}
%\subfloat[][HMM-24]{
%\includegraphics[width=0.09\textwidth]{HMM-24db}}
%\subfloat[][HMM-25]{
%\includegraphics[width=0.09\textwidth]{HMM-25db}}
%\subfloat[][HMM-26]{
%\includegraphics[width=0.09\textwidth]{HMM-26db}}
%
%\subfloat[][CEM-22]{
%\includegraphics[width=0.09\textwidth]{CEM-22db}}
%\subfloat[][CEM-23]{
%\includegraphics[width=0.09\textwidth]{CEM-23db}}
%\subfloat[][CEM-24]{
%\includegraphics[width=0.09\textwidth]{CEM-24db}}
%\subfloat[][CEM-25]{
%\includegraphics[width=0.09\textwidth]{CEM-25db}}
%\subfloat[][CEM-26]{
%\includegraphics[width=0.09\textwidth]{CEM-26db}}

\caption{Intuitive comparison of recovered examples for  DeepLofargram with different architectures  in different SNRs.}
\label{fig:recovered1}
\end{figure}

\begin{figure}
\centering

\subfloat[HMM-22]{
\includegraphics[width=0.09\textwidth]{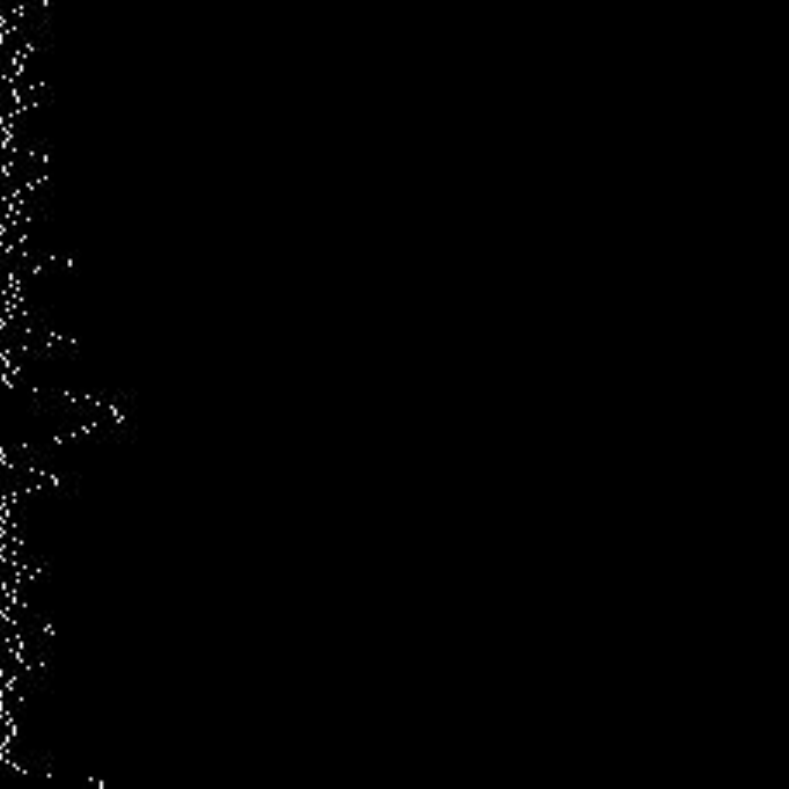}}
\subfloat[][HMM-23]{
\includegraphics[width=0.09\textwidth]{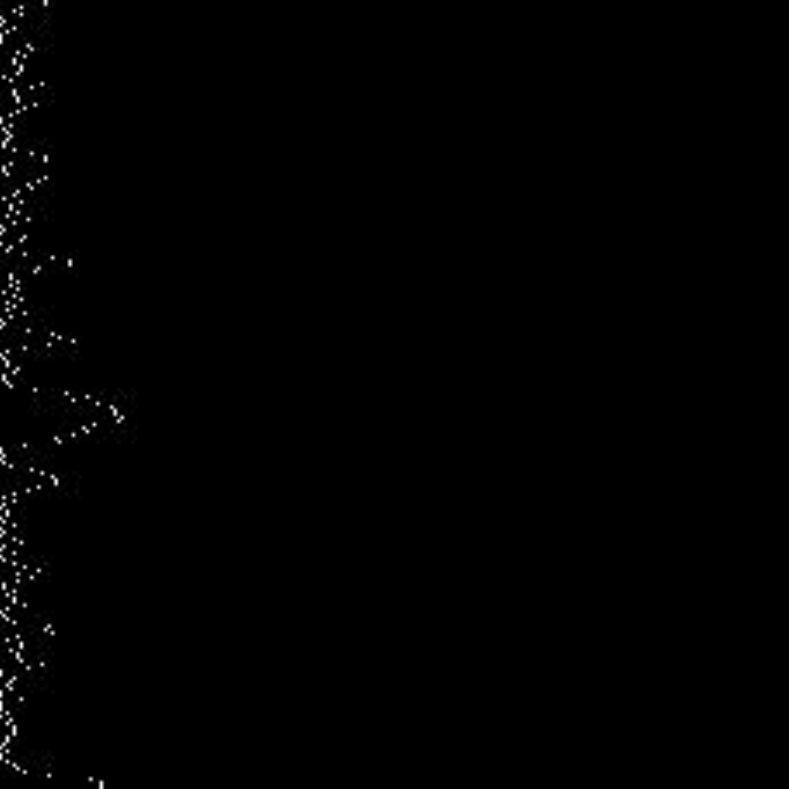}}
\subfloat[][HMM-24]{
\includegraphics[width=0.09\textwidth]{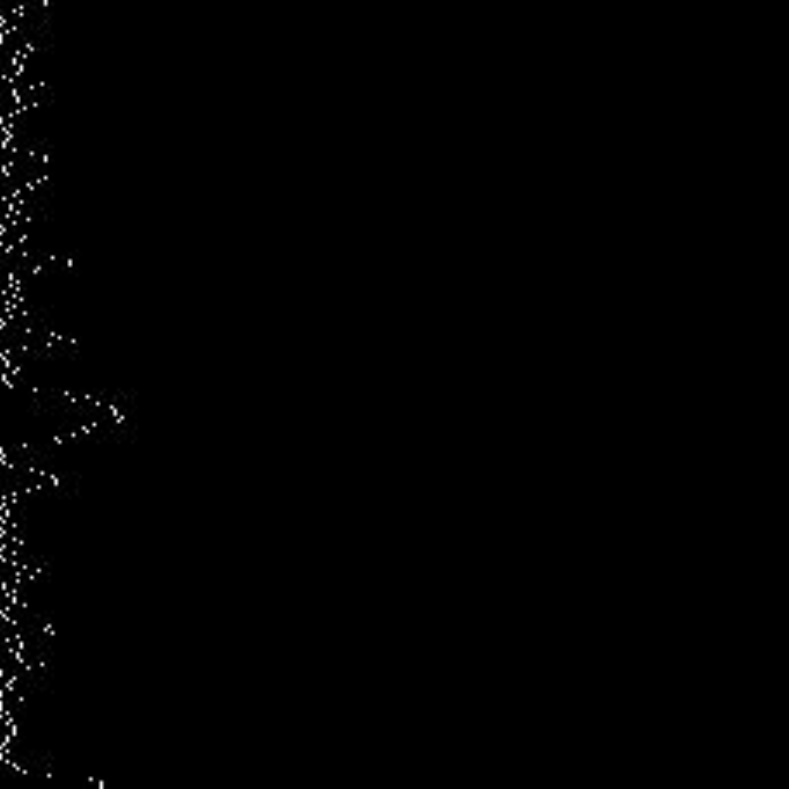}}
\subfloat[][HMM-25]{
\includegraphics[width=0.09\textwidth]{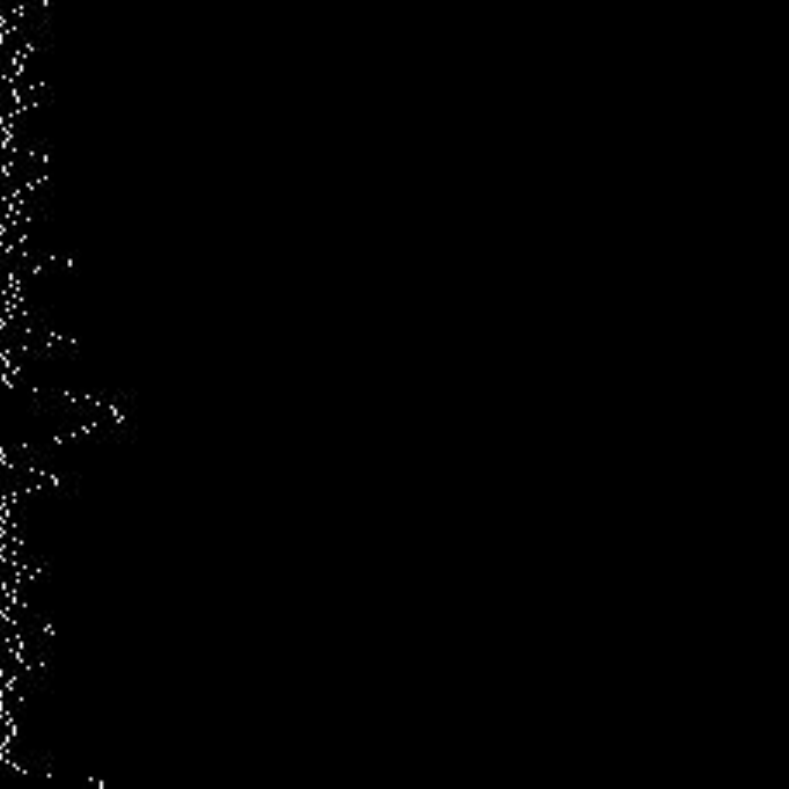}}
\subfloat[][HMM-26]{
\includegraphics[width=0.09\textwidth]{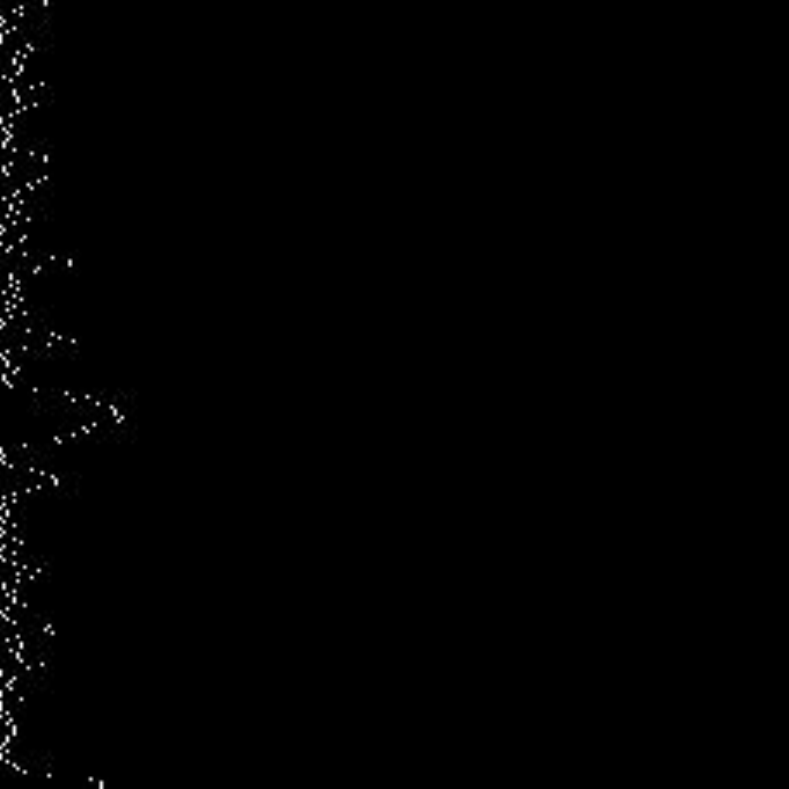}}

\subfloat[][CEM-22]{
\includegraphics[width=0.09\textwidth]{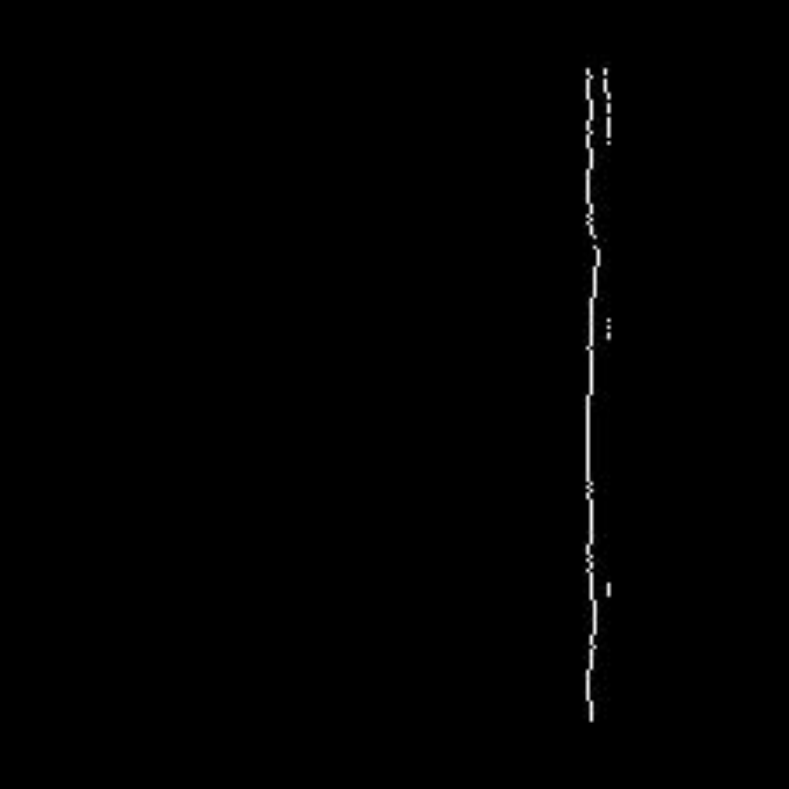}}
\subfloat[][CEM-23]{
\includegraphics[width=0.09\textwidth]{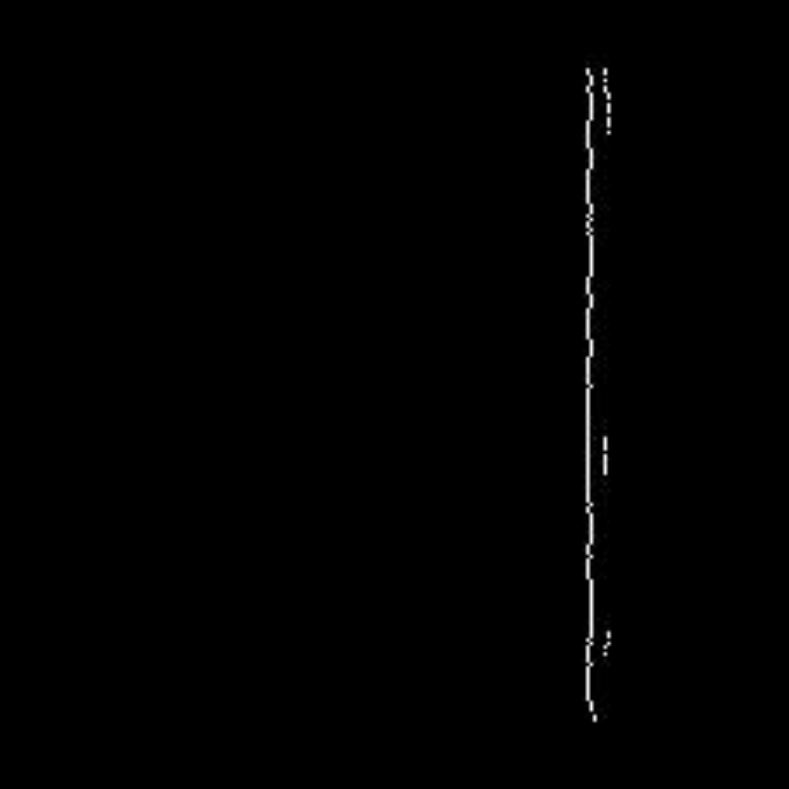}}
\subfloat[][CEM-24]{
\includegraphics[width=0.09\textwidth]{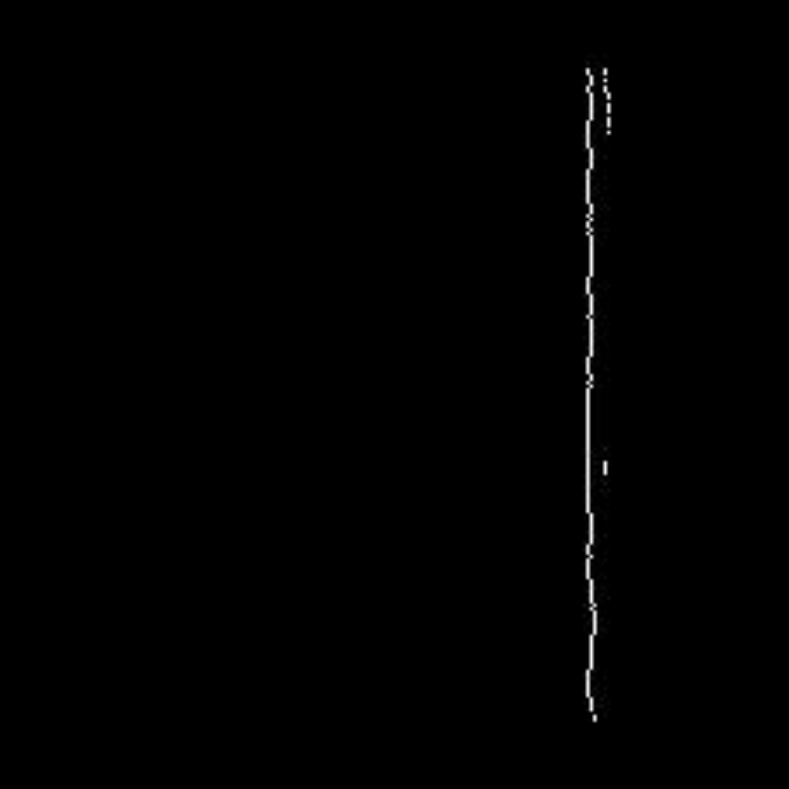}}
\subfloat[][CEM-25]{
\includegraphics[width=0.09\textwidth]{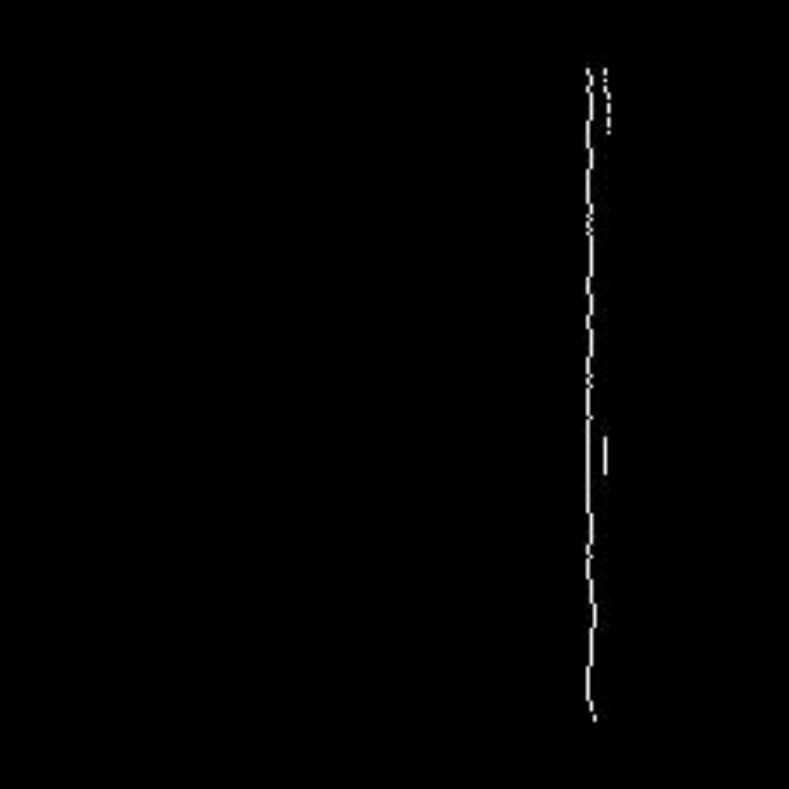}}
\subfloat[][CEM-26]{
\includegraphics[width=0.09\textwidth]{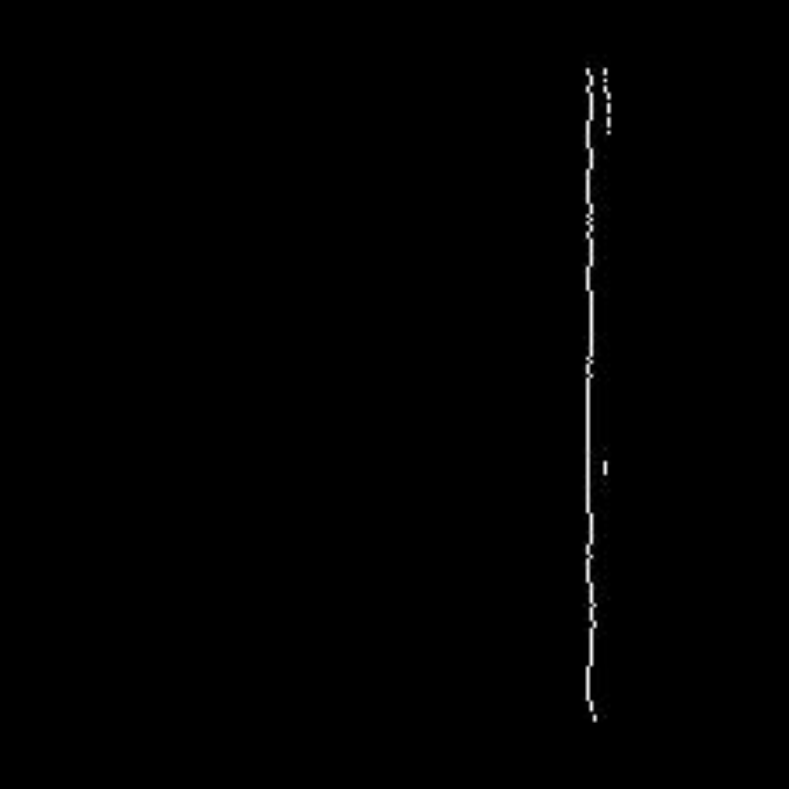}}

\caption{Intuitive comparison of recovered examples for the two state-of-the-art methods in different SNRs.}
\label{fig:recovered2}
\end{figure}

We compare our method with the reported state-of-the-art: contour energy minimisation (CEM)~\cite{ThomasS2013} and HMM-based method~\cite{ParisJ2003}. Nevertheless the focus of these two methods is on frequency line recovery, and we just examine their LLA performances in the same protocol as above.  As shown in Fig.~\ref{fig:lla}, our DeepLofargram outperforms both methods by a large margin. Specifically, for HMM the LLAs are no more than $0.1$, and for CEM the LLAs are constant on the level around $0.14$. Intuitive examples can be found in Fig.~\ref{fig:recovered2}, showing that HMM-based scheme is completely lost in such low SNRs, and what CEM recovered are just the prior input of the base-frequency. DeepLofargram achieves a highly impressive performances, i.e., $0.28\sim0.45$ on average. Intuitive comparisons of recovered examples of Fig.~\ref{fig:dataset} are shown in Fig.~\ref{fig:recovered1}. Satisfactory results can be achieved by DeepLofargram even in the SNR as low as $-26$dB for some.

%\begin{figure}
%\centering
%\subfloat[][Different ConvNet depth]{
%\includegraphics[width=0.24\textwidth]{Depth-lla2}
%\label{fig:depth-lla}}
%\subfloat[][The state-of-the-art]{
%\includegraphics[width=0.24\textwidth]{Methods-lla}
%\label{fig:lla}}
%\caption{LLAs for ConvNets with different depths and the state-of-the-art.}
%\end{figure}

\section{Conclusion}
In this letter we have developed an end-to-end framework, DeepLofargram, to simultaneously detect and recover the fluctuating dim frequency lines submerged in noise. Experiments demonstrate that the performance boundary for DeepLofargram is nearly $-24$dB on average, and $-26$dB for some. This is far beyond the perception of human visual and significantly improves the state-of-the-art performances. Since DeepLofargram leverages on the deep architecture, it can potentially benefit from the abundant emerging new deep neural network techniques and frameworks. Another direction for future research involves engaging DeepLofargram to complex and challenging real-world scenarios such as multiple lines in different SNRs and ocean ambient noise.

\ifCLASSOPTIONcaptionsoff
  \newpage
\fi

\bibliographystyle{IEEEtran}
\bibliography{lofargram2}

\end{document}